\documentclass[namedreferences]{SolarPhysics}
\usepackage[optionalrh]{spr-sola-addons} 
\usepackage{graphicx,color}        
\usepackage{color}           
\usepackage{url}             

\newcommand{\deriv}[2]{\frac{{\mathrm d} #1}{{\mathrm d} #2}}
\newcommand{\tderiv}[2]{{{{\mathrm d} #1}/{{\mathrm d} #2}}}

\renewcommand{\vec}[1]{ {\mathbf #1} }

\newcommand{\bndry}{ {\mathcal S} }

\newcommand{\da}{~{\mathrm d}^2 x}

\newcommand{\bb}{ \vec B}

\newcommand{\uu}{ \vec u}

\newcommand{\xx}{ \vec x}
\newcommand{\ga}{$G_A$}
\newcommand{\gth}{$G_\theta$}

\newcommand{\U}[1]{\ensuremath{\mathrm{~#1}}}
\newcommand{\arcsec}{^{\prime\prime}}

\newcommand{\adv}{    {\it Adv. Space Res.}}

\newcommand{\aap}{    {\it Astron. Astrophys.}}

\newcommand{\ag}{     {\it Ann. Geophys.}}

\newcommand{\apj}{    {\it Astrophys. J.}}

\newcommand{\gafd}{   {\it Geophys. Astrophys. Fluid Dyn.}}
\newcommand{\grl}{    {\it Geophys. Res. Lett.}}

\newcommand{\jastp}{  {\it J. Atmos. Solar-Terr. Phys.}}
\newcommand{\jgr}{    {\it J. Geophys. Res.}}

\newcommand{\nat}{    {\it Nature}}

\newcommand{\solphys}{{\it Solar Phys.}}

\newcommand{\ssr}{    {\it Space Sci. Rev.}}
\newcommand{\hvar}{    {\it Hvar Obs. Bull.}}

\begin{document}

\begin{article}

\begin{opening}

\title{How can a Negative Magnetic Helicity Active Region Generate a Positive
Helicity Magnetic Cloud ? }

\author{R.~\surname{Chandra}\sep
        E.~\surname{Pariat}\sep B.~\surname{Schmieder}\sep C.H.~\surname{Mandrini}\sep
        W.~\surname{Uddin}\sep
       }
\runningauthor{Chandra et al.}
\runningtitle{How can a negative helicity AR generated a positive
helicity MC ?}

\institute{R. Chandra \sep B. Schmieder\\
Observatoire de Paris, LESIA, UMR8109 (CNRS), F-92195, Meudon Principal
Cedex, France\\
email: \url{chandra.ramesh@obspm.fr}, \url{brigitte.schmieder@obspm.fr}\\
\medskip
E. Pariat\\
NASA-GSFC, Space Weather Laboratory, 8800 Greenbelt Rd, Greenbelt, MD 20771,
USA \\
College of Science, George Mason University, 4400 University Dr.,
Fairfax VA, 22030, USA \\
email: \url{epariat@helio.gsfc.nasa.gov} \\
\medskip
C.H. Mandrini\\
Instituto de Astronom\'\i a y F\'\i sica del Espacio (IAFE), CONICET-UBA, Buenos
Aires, Argentina \\
email: \url{mandrini@iafe.uba.ar} \\
\medskip
W. Uddin\\
Aryabhatta Research Institute of Observational Sciences (ARIES), Nainital - 263 129, India
                     email: \url{wahab@aries.ernet.in} \\
             }

\begin{abstract}
The geoeffective magnetic cloud (MC) of 20 November 2003, has been
associated to the 18 November 2003, solar active events in previous
studies. In some of these, it was estimated that the magnetic
helicity carried by the MC had a positive sign, as well as its solar
source, active region (AR) NOAA 10501. In this paper we
show that the large-scale magnetic field of AR 10501 had a negative
helicity sign. Since coronal mass ejections (CMEs) are one of the
means by which the Sun ejects magnetic helicity excess
into the interplanetary space, the signs of magnetic helicity in the AR and MC
should agree. Therefore, this finding contradicts what is expected
from magnetic helicity conservation. However, using for the first
time correct helicity density maps to determine the spatial
distribution of magnetic helicity injection, we show the existence
of a localized flux of positive helicity in the southern part of AR
10501. We conclude that positive helicity was ejected from this
portion of the AR leading to the observed positive helicity MC.
\end{abstract}

\end{opening}


\section{Introduction}
     \label{S-Intro}

Magnetic helicity globally quantifies the signed amount of twist, writhe, and
shear of the magnetic field in a given volume
(see the review by \inlinecite{Demoulin07} and references therein).
Magnetic helicity plays an important role in magnetohydrodynamics (MHD)
because it is one of the few global quantities which are conserved,
even in resistive MHD on  time scales shorter than the global diffusion
time scale (\opencite{Berger84}).

Coronal mass ejections (CMEs) are expulsions of mass and magnetic
field from the Sun. \inlinecite{Rust94} and \inlinecite{Low96}
pointed out that one of the most important roles of CMEs is to carry
away magnetic helicity from the Sun.
Otherwise, because helicity dissipates very slowly and helicities of
opposite sign are globally injected through the photosphere in each solar
hemisphere without change of sign during consecutive cycles
(\opencite{Berger00}), helicity would accumulate
continuously. A fraction of CMEs can be observed {\it in
situ} as magnetic clouds (MCs). An MC is characterized by lower
proton temperature and higher magnetic field strength than the
surrounding solar wind.  Typically, the magnetic field vector shows
a smooth and significant rotation across the cloud
(\opencite{Burlaga81}; \opencite{Klein82}) indicating a helical
(flux rope) magnetic structure, which clearly has non-zero helicity.
Therefore, a measurable prediction is that the interplanetary MC
must carry the same amount of helicity that was ejected from the
solar source region. In particular, the signs of the magnetic helicity of the MC
and of the solar region from which it originates should agree.

  As a first approach, the magnetic helicity sign of some structures in the
solar atmosphere can be inferred from certain observed morphological
features (see the review by \inlinecite{Demoulin09} and references
therein). These features include {\it sunspot whorls} (handedness of
the spiral patterns of chromospheric fibrils), {\it filament barbs}
(direction of the barbs relative to the orientation of the
magnetic field), {\it flare-ribbons} (forward/reverse `J-shape'
observed in H$\alpha$ and UV wavelengths), {\it sigmoids} (normal or
reverse `S-shaped' loops in soft X-ray observations), {\it magnetic
tongues} (angle formed by the magnetic inversion line relative to
the AR axis in emerging ARs), {\it coronal loops} (orientation of
loops in EUV observations relative to the magnetic inversion line),
{\it vector magnetic field} (direction of sheared fields).
These observational features can be used to qualitatively compare
the magnetic helicity sign of an MC with that of its solar source
once identified ({\it e.g.} \opencite{Subramanian01};
\opencite{Schmieder05}; \opencite{Ali07}). 
In a similar way, {\it i.e.} directly from observations, the helicity sign of MCs can be
estimated from the measured rotation of the vector magnetic field
(see examples in Bothmer and Schwenn (1994, 1998)).

Several studies have found that the magnetic helicity sign of
the MC and that inferred from the morphological features of its source AR match 
({\it e.g.} \opencite{RK94}; Bothmer and Schwenn, 1994, 1998; \opencite{Marubashi97};
\opencite{Ruzmaikin03}; \opencite{Rust05}). Other studies have
determined the helicity sign of the CME source region modeling the
coronal magnetic field ({\it e.g.} \opencite{Yurchyshyn01},
\citeyear{Yurchyshyn06}), in these cases also the source region and
cloud helicity signs were in agreement.
However, the comparison does not yield a complete agreement, since a
few MCs seem to present a different helicity sign from that of their
solar source \cite{Leamon04}.

Quantitative comparisons of the helicity involved
in the solar ejection and that of the associated MC have recently
been possible. In the interplanetary medium, these quantitative comparisons
require either the modeling of {\it in situ} magnetic field observations
(see the reviews by \inlinecite{Dasso05} and \inlinecite{Nakwacki08}) or, in
cases when the impact parameter is small
and considering a local cylindrical geometry, the MC helicity can be
directly quantified from the data \cite{Dasso06}.
In the solar atmosphere, at least two different methods, giving consistent
results (\opencite{Lim07}), allow the estimation of the amount of ejected
helicity. One way is to compute the helicity variation
before and after the ejection of the solar source region using a
coronal field model (\opencite{Green02}; \opencite{Demoulin02};
\opencite{Mandrini05}; \opencite{Regnier05}; \opencite{Luoni05}).
This method requires the knowledge of the magnetic field in the
entire volume. Another way is to measure the helicity
injection, based on a time series of photospheric field
observations. This was initiated by \inlinecite{Chae01b} and
\inlinecite{Chae01}, and was subsequently applied to the study of
CMEs by \inlinecite{Nindos02} and \inlinecite{Nindos03}.
\inlinecite{Pariat05} showed that, although the methods used in
previous works could correctly estimate the total injected flux of
helicity, they incorrectly determined the localized injected flux
and proposed an alternative to properly map the helicity flux
injection. We shall use this corrected method to compute the
magnetic helicity injection in AR 10501.

The large MC observed on 20 November 2003, gave
place to the largest geomagnetic storm of solar cycle 23
\cite{Gopalswamy05}. This MC was associated to the active solar
events that occurred in AR 10501 on 18 November 2003, by several
authors (\opencite{Gopalswamy05}; \opencite{Yurchyshyn05};
\opencite{Mostl08}). The association discussed by
\inlinecite{Gopalswamy05} is mainly based on the
timing between the filament ejections in AR 10501, the appearance of CMEs in the Large
Angle and Spectroscopic Coronagraph (LASCO) white light images (its
height-time plots), and the arrival of the MC to the Advanced
Composition Explorer (ACE) spacecraft and its velocity.
\inlinecite{Yurchyshyn05} also estimated the AR helicity sign
modeling the AR coronal magnetic field and the MC {\it in situ}
data; these latter authors showed that the helicity of the MC and AR
were both positive. However, \inlinecite{Mostl08} discussed that
while the MC helicity sign seems well determined,
the handedness (or helicity sign) of the very extended filament, lying along
different portions of the inversion line within and in the surroundings of the AR, is ambiguous.

In this paper, we revisit the evolution of the activity in AR 10501
along 18 November 2003, in Section~\ref{S-obs}. Then, we discuss
the characteristics of the positive-helicity-carrier MC observed by
ACE on 20 November 2003, and its association with the CMEs
originating from the AR (Section~\ref{Ss-MC}). 
To verify this association, we analyze carefully the morphological
features of the region; in particular, the different segments of filament
material lying along the magnetic inversion line that are observed as forming
a single and very extended filament. We find that all
these observational features, except for one filament
segment, indicate that the sign of the dominant helicity in the AR was negative.
Global magnetic field extrapolations, before any AR activity, agree
with this finding (Section~\ref{S-Morph}). This is in strong contradiction with what is
expected by models of MC formation constrained by the
helicity conservation principle. In view of this result, we analyze
the photospheric magnetic field evolution of the region and we
perform an in-depth analysis of the local helicity flux injection
along 18 November (Section~\ref{S-LocHInj}). We find that a zone at
the south of AR 10501 was the location of positive helicity flux
injection during that day. Finally, we conclude that the ejected
flux rope, observed later as an MC, should be located at this
southern portion of the AR,
and discuss the implications of our finding for CMEs/MCs
triggering models (Section~\ref{S-Concl}).


\section{Observations of AR 10501 Activity} 
\label{S-obs}

\subsection{TEMPORAL EVOLUTION OF THE FLARES}
\label{Ss-tempflares}

\begin{figure} 

\vspace{-0.3\textwidth}    
\centerline{\hspace*{-0.05\textwidth}
               \includegraphics[angle=-90,width=2.1\textwidth,clip=]{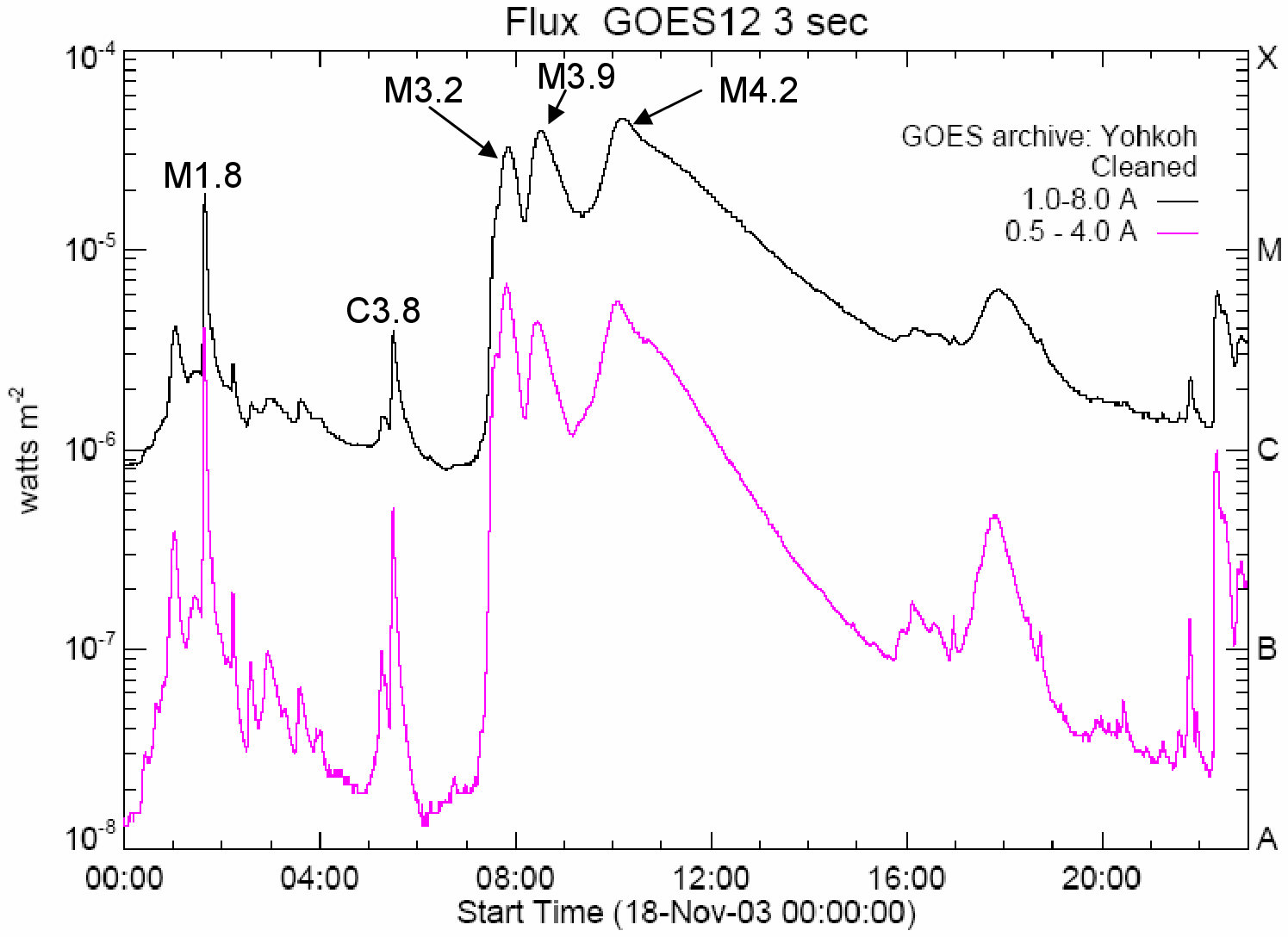}
              }
\vspace{-0.4\textwidth}    
\caption{Temporal evolution of the X-ray solar flux observed by GOES 12
in the 0.5-4 \AA\ and
1-8 \AA\  bandwidths on 18 November 2003.}
   \label{goes}
   \end{figure}

The decay phase of solar cycle 23 has been marked with an unexpected
extreme level of activity, the so-called Halloween events. Figure
\ref{goes} shows the temporal evolution of the flares observed by
GOES on 18 November 2003. Five flares occurred on this day. Three of
them, {\it viz.} the C3.8/SF at 05:25 UT, M3.2/2N at 07:52 UT, and
M3.9/2N at 08:30 UT flares (in terms of X-ray/H$\alpha$ classes) were observed
to originate from AR 10501 (located at N03E08). These flares will be hereafter
referred to as the first, the second and the third flares, respectively. 
These flares were associated with filament eruptions.

The M3.2 and M3.9 flares have been associated with two CMEs,
detected by LASCO \cite{Brueckner95}. The first CME was detected in the
C2 field-of-view at 08:06 UT, had a speed of $\approx 1223
\U{km~s^{-1}}$, and a width of 104$^\circ$. The second halo CME was
observed at 08:50 UT with a speed of $\approx 1660 \U{km~s^{-1}}$ (see
$http://cdaw.gsfc.nasa.gov/CME_{list}/$).

 \begin{figure}    
     \vspace{-0.75\textwidth}    
\centerline{\hspace*{0.07\textwidth}
               \includegraphics[width=1.8\textwidth,clip=]{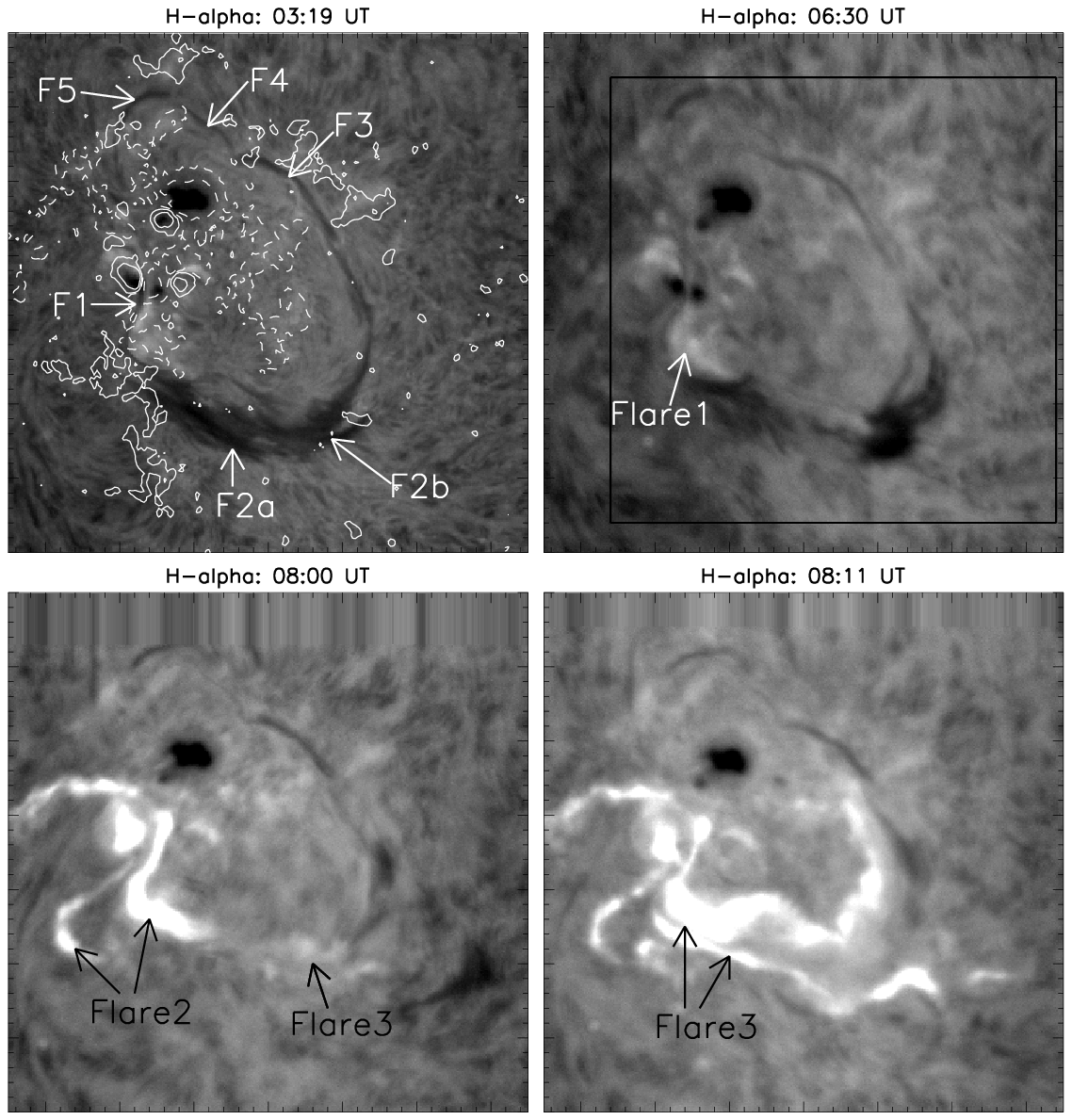}
              }     \vspace{-0.75\textwidth}    

\caption {H$\alpha$ images before (top left panel), after the first 
(top right panel ), and during second (bottom left panel), and third
(bottom right panel) flare. The field-of-view of the all the images
is 350$''\times350''$. Top left: H$\alpha$ image overlaid by MDI
magnetic field contours. The $\pm$100 and $\pm$500 G (gauss) isocontours of
the positive and negative line-of-sight magnetic fields are
plotted with solid and dashed lines, respectively. The arrows indicate the
different filament segments. The arrows in the other panels point to
the three flare ribbons. The square on the top right panel shows the
field-of-view of the images presented in Figure~\ref{trace}. The
arrow in the bottom left panel points to the two faint brightenings
which correspond to the onset of the third flare.}
   \label{halpha}
   \end{figure}

 \begin{figure}    
     \vspace{-0.5\textwidth}    
\centerline{\hspace*{0.07\textwidth}
               \includegraphics[width=1.8\textwidth,clip=]{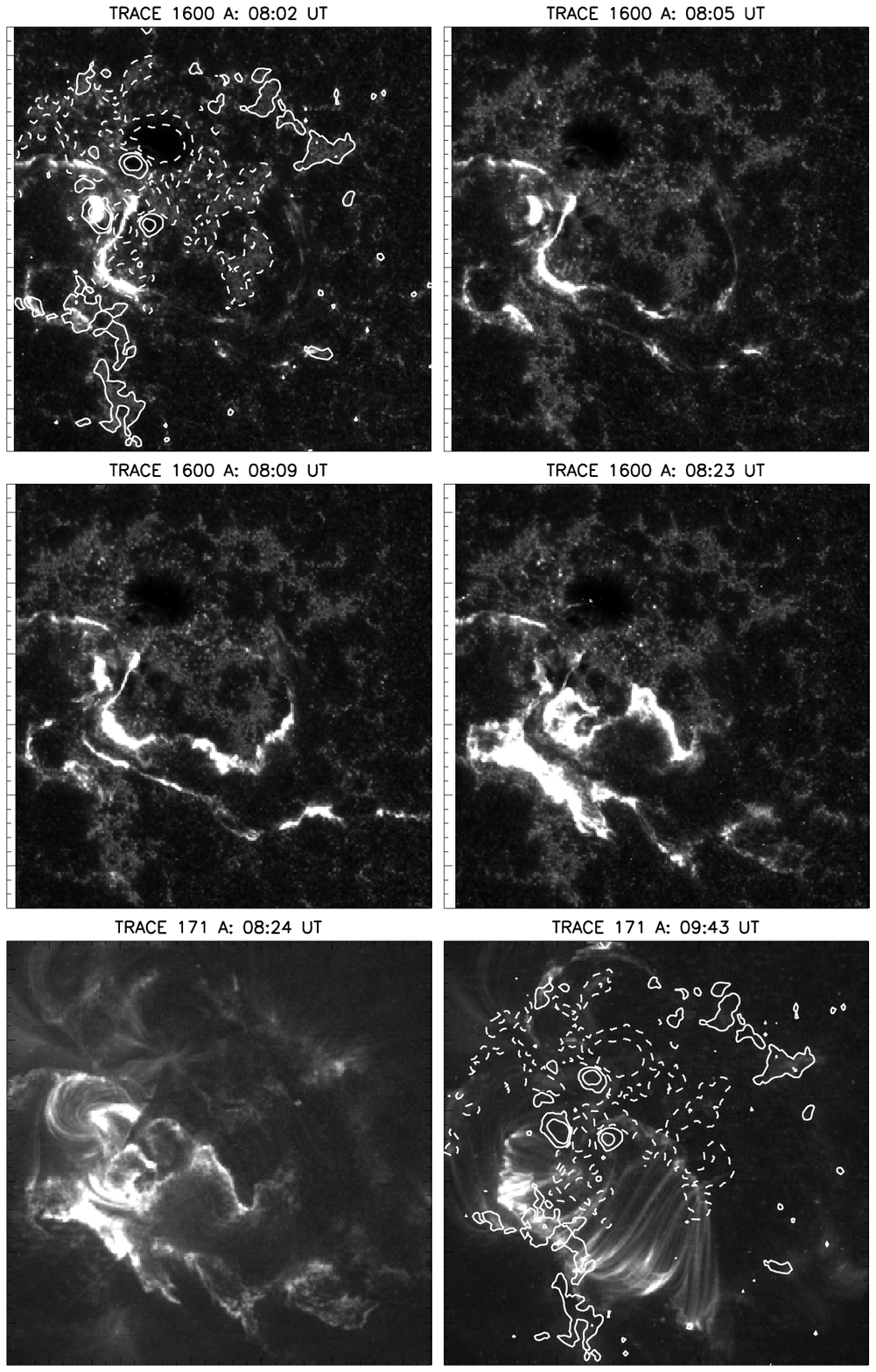}
              }     \vspace{-0.5\textwidth}    

\caption{Upper and middle panels: TRACE 1600 \AA\ images of the
second and third flares. The upper (left) image is overlaid by MDI
magnetic field contours (dashed/solid lines correspond to
negative/positive values). Bottom panels: TRACE 171 \AA\ images of
the third flare during the maximum and decay phase. The right image
is overlaid by MDI contours (dashed/solid lines correspond to
negative/positive values). The isocontours of the line-of-sight
magnetic field are $\pm$100 and $\pm$500 G. The field-of-view of the
images is 300$''\times300''$.}
   \label{trace}
   \end{figure}


\subsection{\bf FILAMENT ERUPTIONS} 
  \label{Ss-Flares}

For this work, we used the H$\alpha$ data from the Aryabhatta Research Institute of
Observational Sciences (ARIES), Nainital, India, obtained with the
15-cm f/15 coud\'e tower telescope equipped with an H$\alpha$ Halle
filter having a pixel size of $1\arcsec$. We also used magnetograms
acquired with the Michelson Doppler Imager (MDI on board the {\it
Solar and Heliospheric Observatory} (SOHO), time cadence of 96
minutes and pixel size of $1.98\arcsec$; \opencite{Scherrer95}) and
171 \AA\ /1600 \AA\ images obtained with the {\it Transition Region
and Coronal Explorer} (TRACE, pixel size of $0.5\arcsec$; \opencite{Handy99}).

All images, H$\alpha$, TRACE, and MDI magnetograms, were co-aligned
to compare the different observed features. Figure~\ref{halpha}
shows H$\alpha$ images before (top left image overlaid with MDI
line of sight isocontours on 18 November 2003 at 06:23 UT)
and during the three flares mentioned above. This AR is formed by a large negative-polarity spot, surrounded 
by weaker positive  polarities. Along the inversion line different segments of
filament material encircle the main negative polarity forming
an apparent large circular-shaped single filament. 
The circular-shaped filaments have been also studied in previous studies (\opencite{Rompolt90}; \opencite{Schmieder07}).
Indeed, a filament may consist of different sections or segments in complex ARs
\cite{Deng02}. Filament models have shown that several magnetic flux tubes sometimes 
have to be
introduced to reproduce the observed distribution of filament material (\opencite{bobra08};
\opencite{Dudik08}). The different segments of the present filament
have been identified and numbered as follows (see
Figure~\ref{halpha}, top left panel): F1 is in the east, F2 with a gap between 
F2a and F2b in the south, F3 in the north-west, F4 and F5 in the
north. The H$\alpha$ image at 06:30 UT (Figure~\ref{halpha}, top
right) shows the segment F1 lying between two small spots at the
south of the main spot, F2a has a large extension towards the east
and F2b a dark extension  towards the west. 
Actually, F2a and F2b may be two separate filaments. This decomposition of F2 
in two segments is also
consistent with the different chirality of each section, as detailed in Section 3.
Nevertheless, we keep the F2 notation for both segments since  the
H$\alpha$ movie reveals counter-streaming motion along F2, some plasma traveling in the east-west direction several times during the pre-eruptive phase.
Counter-streaming in such a huge filament is a signature of
instability \cite{Schmieder08}. F3, F4 and F5 are visible as faint
spines. At 08:00 UT (Figure~\ref{halpha}, bottom left panel),
after the first and second flares, F1 and F2a are no more visible and 
F2b is erupting
(black structure in the right part of the frame).
Simultaneously, bright flare ribbons develop on both sides of
the filament channel of F1. Weaker brightenings then appear along
the filament channel of F2a and  F2b.
This location is the place of the onset  of the third flare
(Figure~\ref{halpha}, bottom right panel). These ribbons are
well observed in TRACE 1600 \AA\ images (Figure~\ref{trace}). The
ribbons expand away from the polarity inversion line as the flare
progresses. The flare ribbons of the third flare develop
towards the east and join the flare ribbons of the second flare
to make a unique system of two ribbons. Figure~\ref{trace} (bottom
right) presents a TRACE 171 \AA\ image of the  post-flare loops
overlaid by MDI magnetic field contours. The filament segments  F3,
F4, F5 are not affected by the different flares, the eruption of F2 and the disappearance of F1.


\subsection{ASSOCIATED MAGNETIC CLOUD ON 20 NOVEMBER, 2003} 
\label{Ss-MC}

The interplanetary cloud associated to the 18 November 2003, solar
active events lasted from 11:16 UT to 18:44 UT on 20 November 2003 when
observed at ACE, according to \inlinecite{Mostl08}. In order to
identify which of the filament ejections could be at the origin of
this MC, we have proceeded as suggested by \inlinecite{Demoulin08}
in his review. He pointed out that, when associating solar
to interplanetary events using their observation times, it is more
reliable to start from the time at which the MC is observed and to
extrapolate backward to the solar surface because, first, {\it in
situ} radial velocities are more precise and, second, these
velocities are closer to the mean velocity during the
travel time of the cloud than the plane-of-sky velocities measured using coronagraph
data.

From ACE data, the MC mean radial velocity was $\approx$ 600
km~s$^{-1}$. If we extrapolate it back to the solar surface, {\it
i.e.} 1 AU distance, the solar ejection should have occurred 69
hours before. Taking $\approx$ 15:00 UT as the time for the MC center
passing across ACE, the solar source event should have occurred at
$\approx$ 18:00 UT on  17 November. The solar source can be searched in
a time window as long as $\pm$ 1 day, as used in some studies
(\opencite{Marubashi97}; \opencite{Watari01}). Then, we searched
from 16 November, 18:00 UT to 18 November, 18:00 UT on the Sun. We
found four CMEs on 17 November ({\it i.e.} at 08:50 UT, 09:26 UT,
13:50 UT, and 23:50 UT) and four CMEs on 18 November ({\it i.e.} at
05:26 UT, 08:06 UT, 08:50 UT, and 09:50 UT). It is generally
believed that an MC will be observed in the Earth vicinity, if its
associated CME is a halo or a partial halo and if its source region
is close to the solar disk center. Therefore, as also proposed by
\inlinecite{Gopalswamy05}, the most probable solar source for the MC
was the CME of 18 November 2003, at 08:50 UT. This is a halo CME
that was associated with the largest flare and eruption from AR
10501, located close the solar disk center at that time.

Different aspects of the MC of 20 November 2003, were studied by
several authors (\opencite{Yurchyshyn05}; \opencite{Gopalswamy05};
\opencite{Wang06}; \opencite{Mostl08}); all of them concluded that
the MC helicity was positive and that its source region was AR
10501. In particular, its magnetic helicity sign can be directly
determined from observations (see Figure 7 in
\opencite{Yurchyshyn05}), {\it i.e.} from the magnetic field
components in the GSE (geocentric solar ecliptic) system of
coordinates it can be seen that this is a ESW cloud (see
\opencite{Bothmer94}), indicating a positive magnetic helicity.


\section{Helicity Deduced from Observed Morphological Features of NOAA AR 10501} 
  \label{S-Morph}

\begin{figure}[h] 

\centerline{\hspace*{0.02\textwidth}
               \includegraphics[width=0.80\textwidth,clip=]{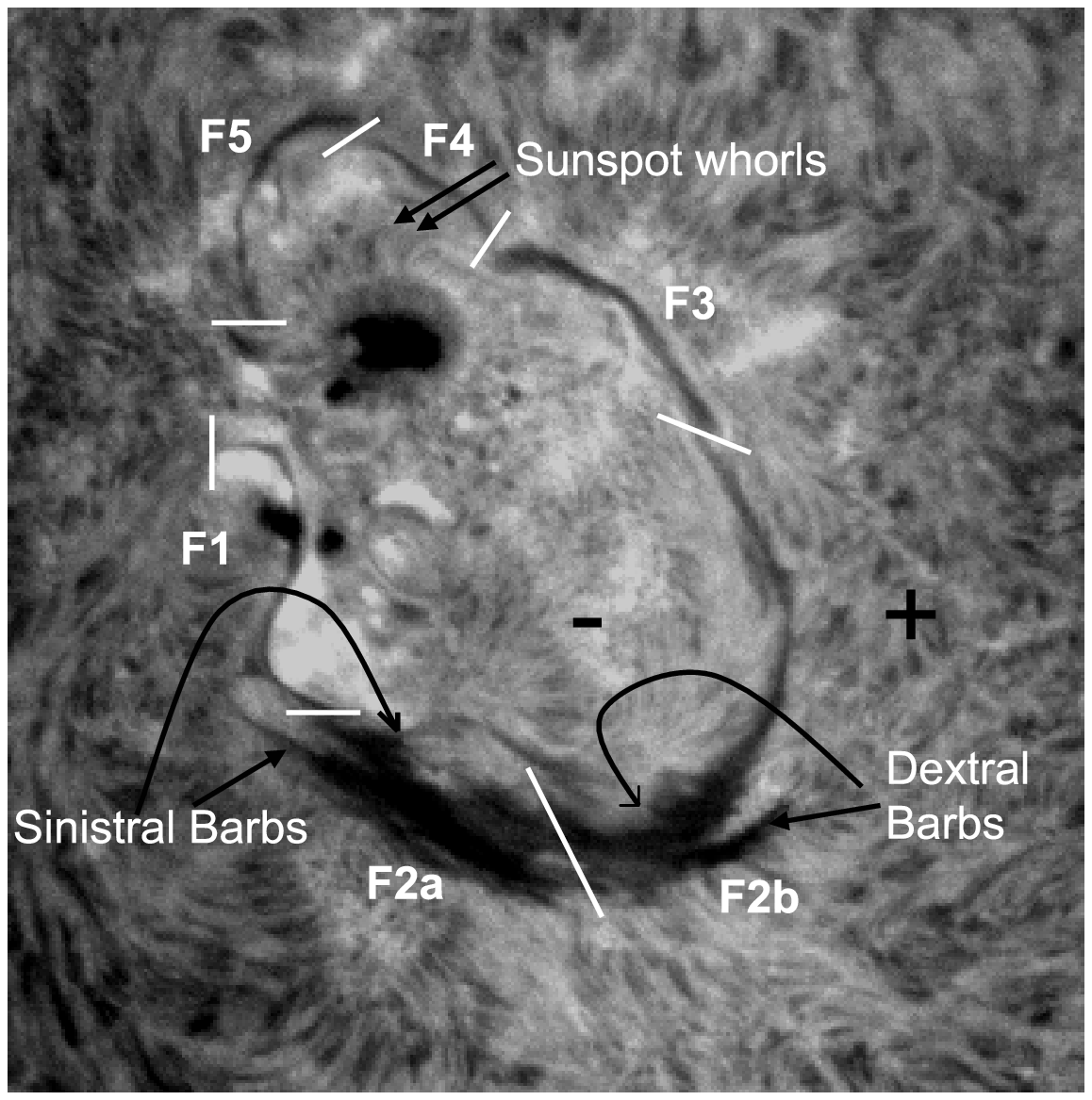}
              }
\caption{ H$\alpha$ image from ARIES (Nainital, India) of AR 10501
on 18 November 2003, at 03:19 UT, before the flares and
eruptions of filament material (the + and - signs indicate the
positive and negative polarity regions. The white bars indicate the
ends of the different segments of the large-scale circular
filament: F1, F2a, F2b, F3, F4, F5).}
   \label{chirality}
   \end{figure}

As discussed in Section~\ref{S-Intro}, the magnetic helicity
sign of an AR can be deduced from several observed morphological
features (\opencite{Demoulin09}). Figure~\ref{chirality} shows an
H$\alpha$ image of AR 10501 illustrating the structures from which
the helicity sign can be inferred.

\begin{enumerate}
\item {\it Sunspot Whorls:} \\
The curvature of the fibrils around sunspots can be used as a tracer
of the helicity sign in an active region (\opencite{Nakagawa71};
\opencite{RustMartin94}; \opencite{Chae01}), since the direction of
the fibrils is related to the twist of the sunspot magnetic field. A
sunspot in which the fibrils turn counterclockwise (clockwise)
towards it will have negative (positive) magnetic helicity, respectively. In
AR 10501, the fibrils around the main spot appeared as
counterclockwise whorls. This indicates that negative magnetic
helicity is carried by the
main spot. \\

\item {\it Shape of Flare Ribbons:} \\

The double forward/reverse `J-shaped' ribbons and their orientation and
displacement along the polarity inversion line indicate the magnetic
helicity sign of the region (\opencite{Demoulin96};
\opencite{Pevtsov96}). In \inlinecite{Chandra09} `J-shaped' ribbons
were associated with emerging flux of positive magnetic helicity. If
we look at Figure \ref{trace} (top left), we find  reverse
`J-shaped' flare ribbons. This could indicate that the AR
globally has a negative magnetic helicity. However, the reverse
`J-shaped' flare ribbons follow the magnetic polarity inversion
line. Therefore, flare ribbons might be a false indicator.\\

\item {\it Filament Segments and Barbs:} \\

\inlinecite{Gopalswamy05} proposed a method to determine the
magnetic helicity based on the sign of the magnetic  polarities of
both filament ends and on the direction of the field within the
filament. Considering one end in the positive polarity (north end of
F3) and the other end in the negative polarity (south end of F1), they
derived a positive sign for the magnetic helicity of the filament material between these
ends. \inlinecite{Mostl08} discussed the choice of the ends of the
eruptive filament material and concluded that a negative
helicity would be the correct one for the filament if
we consider two filaments and not a unique circular filament.
\smallskip

The chirality of a filament can be also directly inferred from
the relative orientation of the barbs, the fine structures along the
filament spine (\opencite{Zirker97}; \opencite{Martin98}).
\inlinecite{Tandberg94} describes this method as follows:
assuming that the filament spine corresponds to a ``high-speed
highway in the United States'', the barbs  of a dextral (sinistral) filament appear as a right (left) 
exit, respectively. In
simplest cases the filament has in general the same chirality as the
associated AR \cite{Lim09}.
However, filaments may not be simple magnetic structures ($\it cf.$
Section~\ref{S-Intro}), and moreover, each segment of a
filament may possess a different chirality, as it has been already
observed by \inlinecite{Martin94} and \inlinecite{Schmieder04} and
modeled by \inlinecite{Devore05}. These works demonstrate that
the segments could not really merge if they have
different chiralities unlike segments having the  same chirality.
\smallskip

A filament in which the barbs, when viewed by an observer on the
positive magnetic polarity side of the filament, are directed
rightward (leftward) is called dextral (sinistral) and  has a
negative (positive) magnetic helicity, respectively. The overlying flux tube
arcades, which are anchored in  enhanced magnetic polarities on both
sides of the magnetic inversion line are left bearing (right
bearing) in case of negative (positive) helicity, respectively \cite{Martin98}.
\smallskip

In the present active region, F1
has no visible barbs while F2  barbs are well identifiable (see Figure~\ref{chirality}). In the
eastern (F2a) part of this filament segment the barbs seem to
correspond to a sinistral filament, while its western part (F2b) to
a dextral filament. The filament segment F2 would thus have two
sections of opposite magnetic helicity. This determination remains
again relatively ambiguous. To confirm these observations  other
methods - extrapolation and computation of magnetic helicity- need
to be used (see Sections 4, 5 and 6). The filament segment F2,
where counter-streaming occurs could be modeled by a flux tube with two
different sections having the same central axis direction, but with opposite twists
in  F2a and F2b.

\end{enumerate}


 \begin{figure}[t]    

 \centerline{\hspace*{0.010\textwidth}
              \includegraphics[width=1.00\textwidth,clip=]{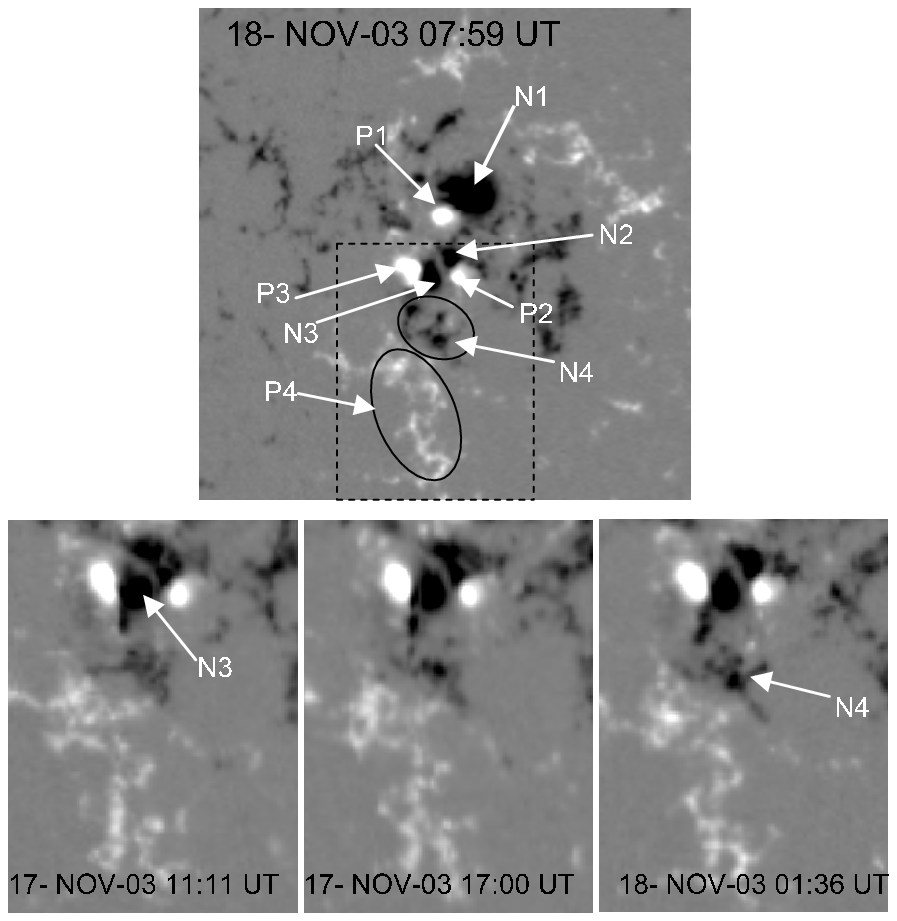}
}
\vspace*{-1.8cm} \caption{Top panel: SOHO/MDI magnetogram of AR
10501 on 18 November 2003, before flare onset. The different
polarities are pointed by arrows (P1, P2, P3, P4: positive polarities
and N1, N2, N3, N4: negative polarities). Bottom panel: enlarged view
of the rectangular area on the top panel showing the evolution of
parasitic polarities.}
   \label{mdi}
   \end{figure}


\begin{figure}   
\centerline{\hspace*{0.100\textwidth}
               \includegraphics[width=0.60\textwidth,viewport=25 130 587 600,clip=]{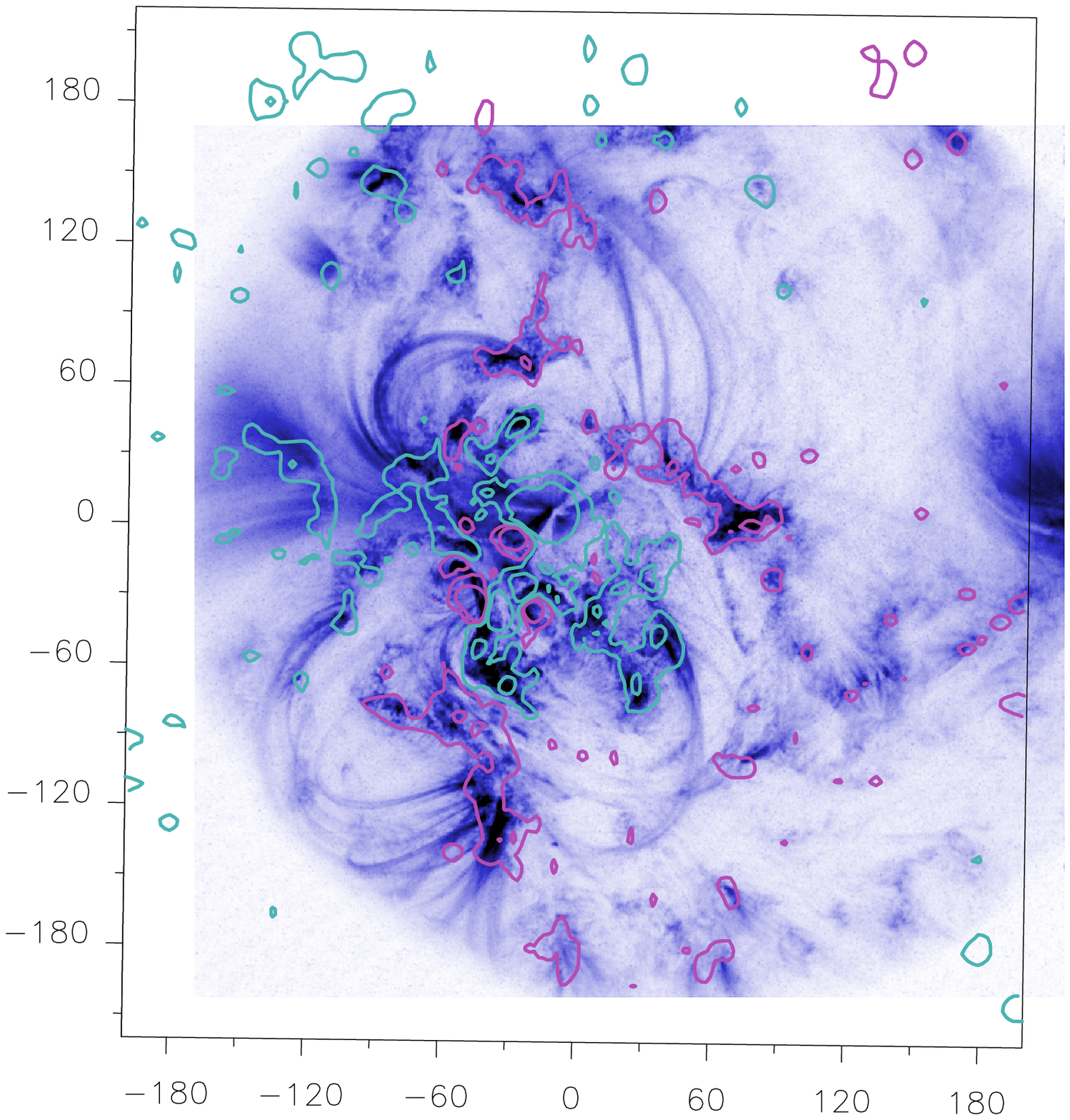}
               \hspace*{-0.08\textwidth}
               \includegraphics[width=0.60\textwidth,viewport=25 130 587 600,clip=]{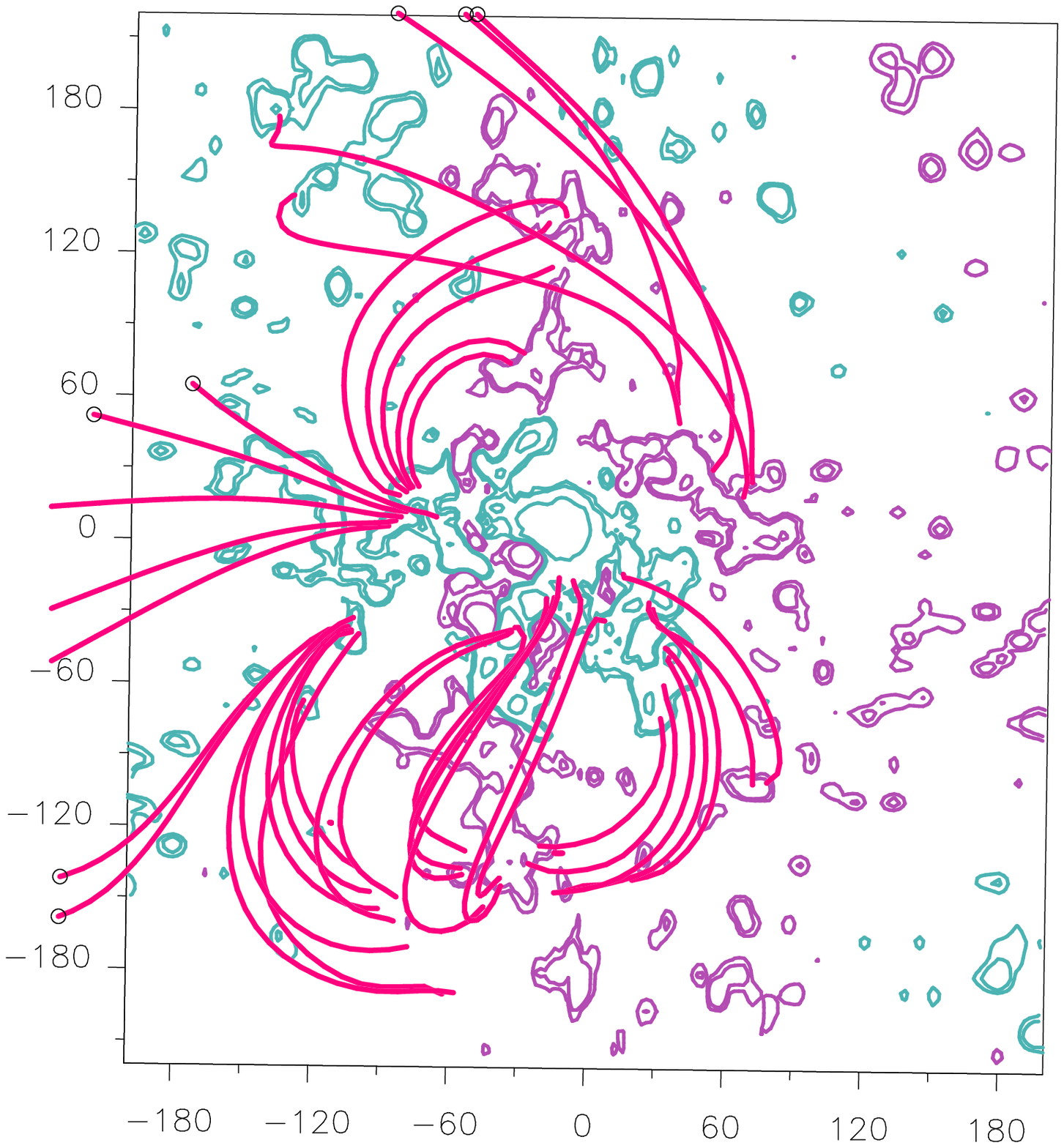}
              }
     \vspace{-0.35\textwidth}   
     \centerline{\Large \bf     
         \hfill}
     \vspace{0.35\textwidth}    
\vspace*{-0.5cm}
\caption{TRACE (171 \AA, 05:24 UT) overlaid by
SOHO/MDI contours (at 06:23 UT) (left) and field lines tracing the
observed loops (right), computed from linear force-free-field models with
different values of the $\alpha$ parameter (see text for details).
The contours correspond to $\pm$50, $\pm$100, and
$\pm$500 G ( red for positive and cyan for negative polarities).}
   \label{mdi1}
   \end{figure}

\section{Evolution and Model of the Magnetic Field in AR 10501} 
  \label{S-Topology}

The magnetic field evolution of AR 10501 before the flares and
filament eruptions, on 17 and 18 November, is presented in
Figure~\ref{mdi}. AR 10501 is a decaying active region, the return
of the very flare productive AR 10484 during October and November,
2003. The AR is formed by several magnetic polarities, {\it viz.}
P1, P2, P3, P4 (positive polarities) and N1, N2, N3, N4
(negative polarities), as marked in the upper panel of
Figure~\ref{mdi}. In the period preceding the events studied in
this paper we observe the displacement of several small positive and
negative polarities towards the south of the AR. These polarities
break from the main AR spot and go away from it. These are the
so-called moving magnetic features, observed typically during the
decay stage of spots as they diffuse. The small polarities merge
with others of opposite sign as they moved away from the spot. After
19 November, the negative polarity N2 breaks into two parts and both
broken parts are seen to rotate around P3. On 20 November 2003, P2
and N2, that are fading, are no longer visible.

We have extrapolated the photospheric magnetic field to the corona
before the flares using a linear force-free-field approximation
($\vec{\nabla} \times \vec{B}= \alpha \vec{B}$, where $\vec{B}$ is
the magnetic field). We have done this model using as boundary
condition a magnetogram at 06:23 UT on 18 November, slightly earlier
than the events discussed in Section \ref{S-obs}. The value of the
free-parameter of the model, $\alpha$, is set by matching the shape
of the computed field lines to that of observed coronal loops, as
discussed in \inlinecite{Green02}. The TRACE image in 171 \AA\ at 05:24
UT, the closest in time to MDI magnetogram, is used to select
$\alpha$. This image allows us to clearly observe the shape of
large-scale coronal loops (see Figure \ref{mdi1}, left panel). The
result of our model, compared to TRACE loops, is shown in Figure
\ref{mdi1}. We have not been able to match all loops using a single
value of $\alpha$ what is expected since, in general, the magnetic
field is not a linear force-free field. Therefore, different sets of loops
were modeled using different values for $\alpha$. However, all the
values needed to find the {\bf } best match turn out to be either
zero (for field lines anchored on the central east weak negative
polarity and going away from the AR) or negative: the values of
$\alpha$ being in the range [-1.2,-0.6]$\times$10$^{-2}$Mm$^{-1}$.
These negative values of $\alpha$ correspond to a negative magnetic
helicity. Therefore, the extrapolation of the magnetic field also
confirms that the large-scale magnetic field of AR 10501 has a
negative helicity sign, probably dominated by the negative
helicity of the main AR bipole (N1/P1) (see Section 6).

\inlinecite{Yurchyshyn05} have modeled the `post-flare' arcade
after the two-ribbon flare (third flare) that peaked at 08:31 UT. In
order to determine the value of $\alpha$, these authors compared
their computed field lines to the half-resolution 195 \AA\ image
obtained by the Extreme-Ultraviolet Imaging Telescope (EIT) at 09:36
UT  (see their Figure 6). \inlinecite{Yurchyshyn05} found that the value of
$\alpha$ that best matched the EIT loops was slightly positive. This
result led them to conclude that the MC and AR helicity signs were
in agreement.

\begin{figure}[h] 

\centerline{\hspace*{0.02\textwidth}
               \includegraphics[width=0.80\textwidth,viewport=25 130 587 600,clip=]{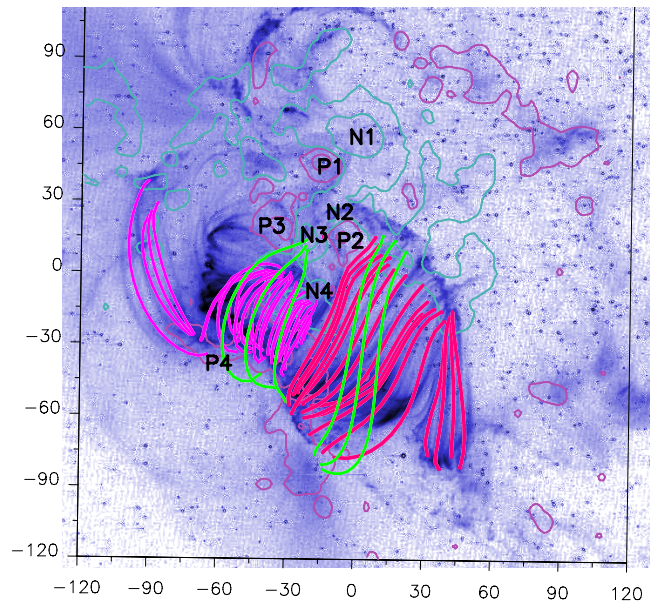}
              }
\caption{The TRACE image (171 \AA, 09:43 UT) overlaid by SOHO/MDI contours (at
09:35 UT) and field lines tracing the observed 'post-flare' arcade
loops. These have been computed using linear force-free models with
different values of the $\alpha$ parameter (see text for details).
The contours correspond to $\pm$100 and $\pm$500 G (red for positive and cyan for negative
polarities).}
\label{arcade-extrap}
\end{figure}

We have also modeled the `post-flare' arcade loops, but in this
case we have compared our computed field lines to the 171 \AA\ TRACE
image at 09:43 UT (see Figure~\ref{trace}, bottom right panel). This
image has a resolution four times better than the EIT image used by
\inlinecite{Yurchyshyn05}, and the arcade loops can be seen clearer. At
the beginning of the flare the arcade loops are anchored at P4 and
N4, as shown by the H$\alpha$ bright ribbons in Figure~\ref{halpha}
bottom panels. However, as the flare progresses and the H$\alpha$
ribbons separate only the eastern portion of the arcade remains
rooted in P4 and N4, probably due to the presence of the strong
polarities P2 and N2, that hinder the ribbon separation (see
Figure~\ref{trace}, bottom right panel). The western portion of the arcade,
as also indicated by the location of the H$\alpha$ ribbons, is still
anchored in P4 while its negative footpoint lies on the weak
field region to the west of P2. Figure~\ref{arcade-extrap} shows the
result of our model overlaid on the TRACE image at 09:43 UT. The
arcade loops to the east, anchored at P4 and N4, are represented by
the set of pink field lines. These field lines have been computed
using either a potential field model or a positive value of $\alpha$
= 6$\times$10$^{-3}$Mm$^{-1}$. We want to remark that we have not
been able to model the loops located to the north of this set of field lines. we 
believe that they were part of the `post-flare' arcade corresponding to the first
flare. On the other hand, the arcade loops to the west are
represented by the set of red field lines; these field lines have
been computed using a negative value of $\alpha$ = -6
$\times$10$^{-3}$Mm$^{-1}$. Since a linear force-free-field model can
effectively represent the observed loops when they have a scale size
of the order of $\alpha^{-1}$ (see \opencite{Demoulin97}), one can
argue that the short eastern loops could have been modeled using
either a low positive, negative, or null $\alpha$ value. To show
that there exists a tendency for a positive value towards the east
and negative towards the west part of the arcade, we also show a set
of larger scale pink field lines anchored at P4 and at a northern
negative polarity region for which $\alpha$ =
6$\times$10$^{-3}$Mm$^{-1}$. Though these field lines do not belong
to the arcade, they clearly show that loops anchored at P4 have a
tendency for positive magnetic helicity. This is in agreement with
our finding of a sinistral filament segment (F2a) towards the east, mainly along the
inversion line between P4 and N4, and a dextral
filament segment (F2b) towards the west, mainly along the inversion
line between weaker field polarities (not numbered in our
Figure~\ref{mdi}). For comparison and to also show the way the arcade
field lines would evolve as the field relaxes to a potential state,
we have added a set of potential ($\alpha$=0) field lines drawn in light green above
both the east and west portions of the `post-flare' arcade. These
results show that the AR magnetic helicity can be locally positive,
as we will also demonstrate in Section~\ref{S-Helicity}.

\section{Photospheric Flux of Magnetic Helicity Flux in AR 10501 } 
\label{S-Helicity}
Recent theoretical developments and improvements in magnetic field
measurements allow us to quantitatively determine the amount of
magnetic helicity injected in an active region (see the review by
\inlinecite{Demoulin09}).

\subsection{METHOD}
\label{Ss-Hmethod}

In order to compute the helicity flux injected in AR 10501, we
determine the proxy for the helicity flux density, \gth, at the
position $\xx$, as suggested by \inlinecite{Pariat05}:
\begin{equation}       \label{Eq-G_theta}
        G_{\theta }(\xx) = -\frac{B_n}{2\pi}
        \int_{S_{\rm p}} \deriv{\theta (\xx - \xx ^\prime )}{t} ~B_{\rm n}^\prime\da^\prime \,.
\end{equation}
in which  $B_{\rm n}$ is the normal component  of the magnetic field
$\bb$, $S_{\rm p}$ is the surface over which the helicity is estimated
(here, the photosphere), and the rotation rate $\tderiv{\theta (\xx
- \xx ^\prime)}{t}$ , of a couple of points $\xx$ and $\xx^\prime$,
is given by:
\begin{equation}
\deriv{\theta (\xx - \xx ^\prime)}{t} =
      \frac{ [ (\xx - \xx ^\prime ) \times (\uu - \uu ^\prime) ]_{\rm n} }
           { |\xx - \xx ^\prime|^2 }  \,.
\end{equation}
The velocity field $\uu$, called ``flux transport velocity'' (\opencite{Welsch06}), corresponds to
the motions of the magnetic structure at the photospheric surface.

A key aspect of the derivation  of the helicity injection is to properly
derive the flux-transport velocity field $\uu$. A larger and larger
number of velocity-inversion techniques now exist to derive the
motion of magnetic features at the solar surface
(\opencite{Welsch07}; \opencite{Chae08}).  Methods based on local
correlation tracking (LCT) (\opencite{November88}) have extensively
been used to derive helicity injection in ARs (${\it e.g.}$
\opencite{Chae01a}; \opencite{Nindos03}; \opencite{Pariat06};
\opencite{Jeong07}). However, \inlinecite{Schuck05} clearly
demonstrated that the local correlation tracking was inconsistent
with the magnetic induction equation. \inlinecite{Schuck06} proposed
an improved method, the differential affine velocity estimator
(DAVE), which uses a variational principle to minimize deviations in
the magnitude of the magnetic induction equation constrained by an
affine velocity profile. DAVE has been particularly efficient to
determine the correct helicity values, in a test case using
numerically simulated data, compared to the more traditional LCT method
(\opencite{Welsch07}).

Even though DAVE4VM, an improved method using vector magnetograms as
input, has been developed \inlinecite{Schuck08}, in the present
paper we used the DAVE algorithm since only line-of-sight
magnetograms are available. In order to determine the velocity flux,
we use the 96-minutes cadence MDI full-disk magnetograms.  We select
a portion of MDI magnetograms with a field of view of $477 \arcsec
\times 457 \arcsec= 346 \U{Mm} \times 332 \U{Mm}$ centered on the
AR. These sections are large enough to assume that the AR is
isolated from other ARs in the helicity computation. We remove solar
differential rotation by taking as a reference time the moment when
the AR was crossing the central meridian. Following
\inlinecite{Labonte07}, we corrected the magnetic field by a factor
of 1.56 (as recommended by \inlinecite{BergerLites03}) and we
multiplied the line-of-sight field by the secant of the angle
corresponding to the location of the center of the AR relative to
the disk center, so as to obtain an approximation of the normal
component of the field at each time. In order to limit the errors
due to strong geometrical deformation induced by the projection of
the AR on the observation plane, we determine the helicity flux only
when the AR was within a radius of $41^\circ$ from solar disk
center. Therefore, we can compute the helicity accumulation only
after 16 November, 11:11 UT, {\it i.e.} two days before the flares.
To derive the velocity field with DAVE we use a window size of
$\omega_0=11$ pixels, as suggested in \inlinecite{Schuck06} for
optimum performance.


\subsection{HELICITY FLUX}
\label{Ss-Hflux}

The traditional diagnostic to estimate the magnetic helicity in an
AR, is to compute the accumulated helicity. The helicity flux
$\tderiv{H}{t}$ can then be simply estimated by integrating \gth\
over $S_{\rm p}$ whereas the helicity accumulation, $\Delta H$ between two
times $t_{\rm i}$ and $t_{\rm f}$ can be determined by the time integration of
the helicity flux:
    \begin{equation}  \label{Eq-dH-theta}
        \left. \deriv{H}{t} \right|_{\bndry_{\rm p}}  = \int_{S_{\rm p}} G_{\theta} \da  \quad \textrm{and} \quad
    \Delta H =\int_{t_{\rm i}}^{t_{\rm f}} \left. \deriv{H}{t} \right|_{\bndry_p} dt
        \end{equation}
with $t_{\rm i}$ and $t_{\rm f}$ the Initial and Final times.

\begin{figure} 

\centerline{\hspace*{0.02\textwidth}
               \includegraphics[width=0.50\textwidth,clip=]{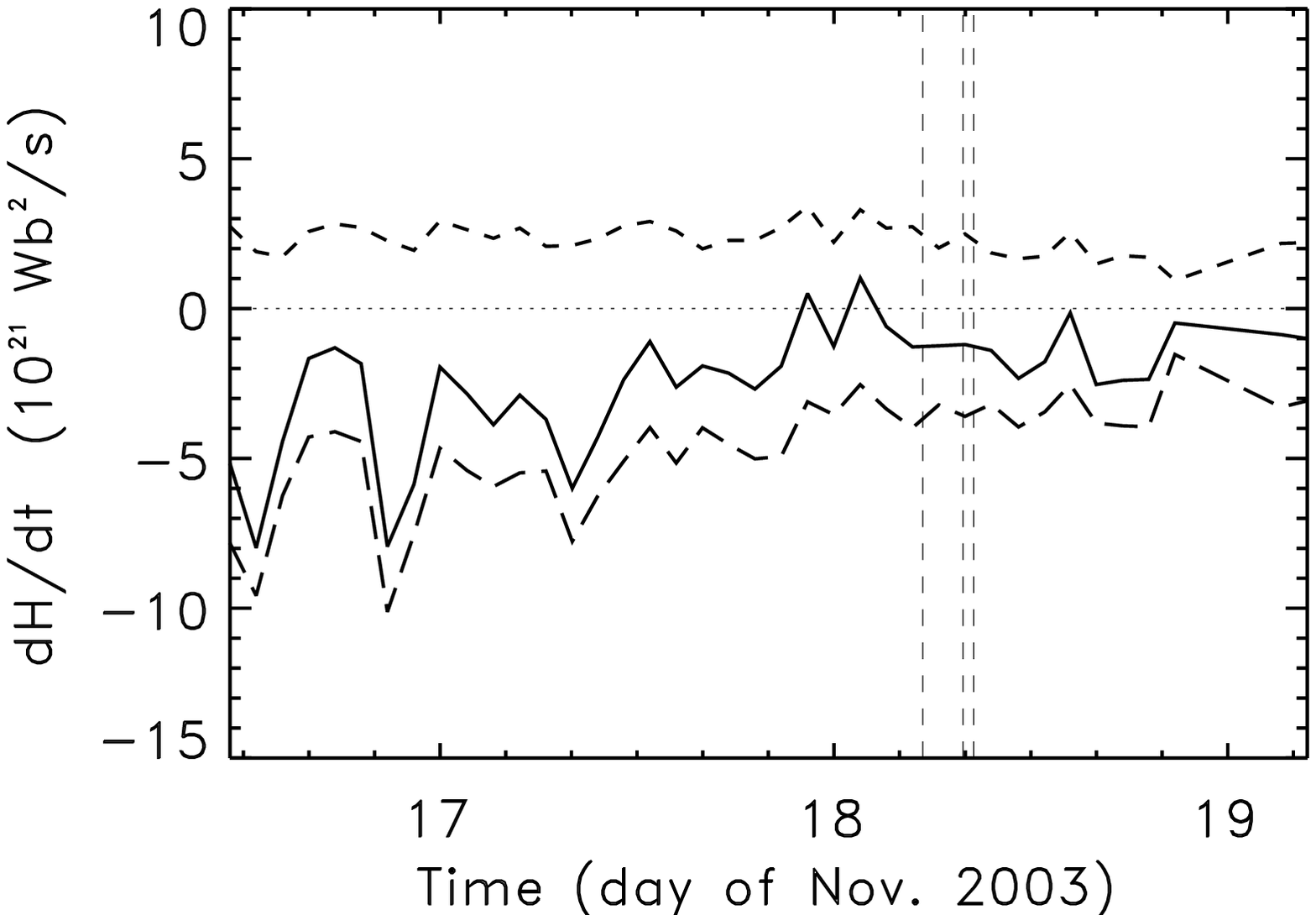}
                \includegraphics[width=0.50\textwidth,clip=]{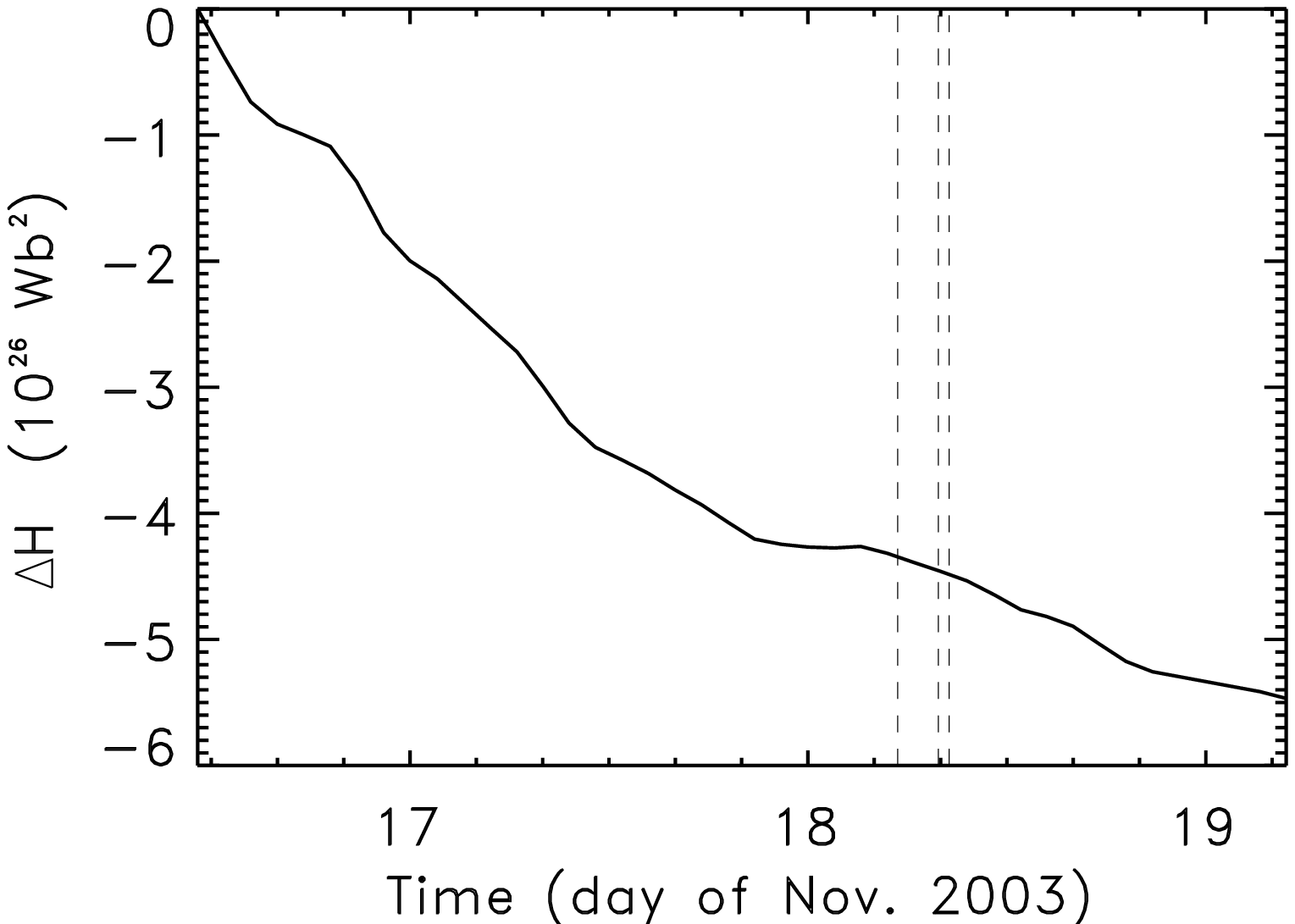}
              }
     \vspace{0.05\textwidth}    
\caption{Signed and total helicity fluxes (left) and accumulated
helicity (right) in AR 10501 between 16 November 11:11 UT and 19
November 04:51 UT.  The vertical dashed lines correspond to the time
of the peak intensity of the three flares described in Section
\ref{S-obs}. In the left panel, the positive helicity flux
(short-dashed line) corresponds to the integrated helicity density
of positive sign while the negative helicity flux
(long-dashed line) corresponds to the integrated helicity density of
negative sign. The total helicity flux (continuous line) is the sum
of the signed fluxes.}
   \label{F-Hflux}
   \end{figure}

In order to control the errors in the estimation of $\uu$, we use
different values of $\omega=$13, 15 and 19 pixels. We find that the
mean variation for the whole helicity computation (while the AR has
an heliocentric angle of $41^\circ$) is about $0.8\times 10^{21}
\U{Wb^2\,s^{-1}}$.  We have also performed helicity computations after
having introduced a random uniform noise of 20 G or 50 G in the
magnetograms. The former is of the order of the noise levels in MDI
measurements while the latter is larger than those (\opencite{Scherrer95}). 
We find that they induce a
mean variation of $0.3\times 10^{21}$ and $0.8\times 10^{21}$ Wb$^2s^{-1}$, respectively, relative 
to the original
data. However, this does not warrant wrong estimations inherent to
the poor temporal resolution (96 minutes!) and to the intrinsic use
of the flux-transport velocity $\bf u$ (${\it e.g.}$ 
\opencite{Demoulin03}; \opencite{Welsch07}; \opencite{Demoulin09}).
Given these uncertainties, the measured helicity flux and helicity
are shown in Figure \ref{F-Hflux}.


The main result of the helicity flux estimation confirms the results
of the topological (Section~\ref{S-Topology}) and morphological
(Section~\ref{S-Morph}) analysis of AR 10501: negative injection of
helicity is dominant.  In the two days preceding the flares, more
than  $-4\times 10^{26} \U{Wb^2}$ ($-4\times10^{42} \U{Mx^2}$) have
been injected in the AR. The helicity accumulation profile presents
two stages, first a relatively constant injection at a rate of the
order of $-8\times 10^{24} \U{Wb^2 hr^{-1}}$, and then a relatively
constant flux a few hours before the eruption. This agrees with
previous studies of intense flaring regions (\opencite{Park08}). AR
10501 is a mature region; therefore, a much larger amount of
helicity has probably been injected before. Most intensely flaring
ARs have a helicity larger than $10^{26} \U{Wb^2}$
(\opencite{Labonte07}) with an average helicity for erupting regions
of about $13\times 10^{26} \U{Wb^2}$ (measured
from the emergence of AR by \inlinecite{Nindos04}, but we corrected an erroneous factor 2
in their Equation (1)). There is, therefore, enough helicity to generate a
CME and an MC for which the average helicity is estimated to be
about $10^{26} \U{Wb^2}$ ({\it e.g.} \opencite{Devore00};
\opencite{Lynch05}). However, the problem remains that the helicity
accumulation of the AR is opposite to the MC in strong
contradiction with what is expected ({\it cf.} Section ~\ref{S-Intro})!

How significant is the positive helicity injection? In
Figure~\ref{F-Hflux} one observes that, although the negative helicity
injection (long-dashed curve) is largely dominant on 16 November, it
seems to decrease by the end of 17 November. The positive injection
is even slightly dominant a few hours before the flares. However,
the total helicity flux values lie within the error margin and are
below the noise threshold. Different choices of $\omega$ with DAVE
sometimes present no dominance of positive helicity. Therefore, it would be a
strong overstatement to sustain that there is dominance of positive helicity
few hours before the flare. We also do not see any particular change
of the sign of the helicity flux during these flares.

Nonetheless, a particular characteristic of the helicity injection
in AR 10501 is that both positive and negative fluxes of helicity tend to be large compared to the
total helicity flux. The injection of positive helicity (short-dashed
curve, Figure~\ref{F-Hflux}, left panel) is relatively constant and
is in the range $2-3 \times 10^{21} \U{Wb^2\,s^{-1}}$. Is this flux
injection due to a localized injection of helicity? If confirmed,
this could result in a helicity accumulation larger than $4\times
10^{26} \U{Wb^2}$, which could be sufficient to explain the
generation of a positive helicity MC.

However, one must be extremely careful when studying  positive and negative helicity
fluxes separately, since the methods estimating the helicity may
produce artifacts. \inlinecite{Pariat05}
demonstrated that the previous definition for the helicity density,
called \ga, requiring the computation of the vector potential of the potential magnetic
field ${\bf A}_{\rm p}$, leads to the generation of strong spurious
signals of opposite helicity sign. \inlinecite{Pariat05}, defined a
new helicity density, \gth, used in the present study, that
significantly reduced the intensity of the fake helicity fluxes.
Applying this new definition to ARs, \inlinecite{Pariat06} showed
that the intensity of the non-dominant fluxes were reduced and that the
helicity injection was much more uniform and unipolar ({\it i.e.} of only one
sign) than previously found. However, \gth, as a proxy for the
helicity flux density, is not exempt of spurious signals
(\opencite{Pariat07}). It is, therefore, necessary to do a careful
analysis of the helicity density map to determine if the local
helicity fluxes are real or not.


 \section{Local Injection of Magnetic Helicity in AR 10501}
 \label{S-LocHInj}

 \begin{figure} 

\centerline{\hspace*{0.02\textwidth}
               \includegraphics[width=0.90\textwidth,clip=]{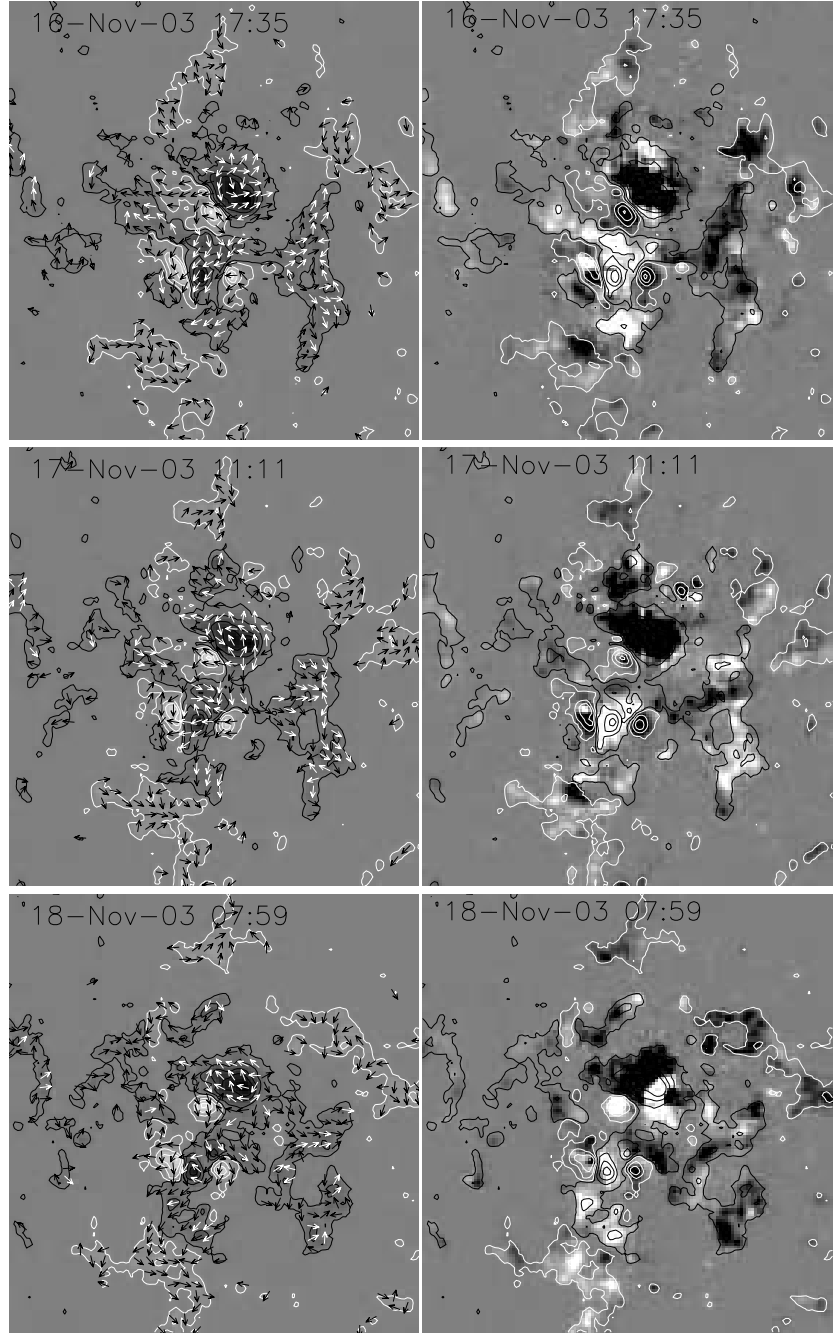}
              }     \vspace{0.05\textwidth}    
\caption{Evolution of the magnetic field and of the helicity
injection in AR 10501 on November 2003. The field-of-view is of $279
\arcsec \times 299 \arcsec= 202 \U{Mm} \times 217 \U{Mm}$. Left
panels: line-of-sight ($B_n$) MDI magnetograms with flux transport
velocity field $\uu$ (arrows). Right panels: helicity flux density
(\gth) maps. Note that the saturation level, equal to $10^{6}
\U{Wb^2\, m^{-2}\,s^{-1}}$, has been chosen to enhance the
distribution of helicity in the magnetic polarities of intermediate
intensity. The contours correspond to 100,
600, 1200, 1800 G (white for positive and black for negative polarities).
}
   \label{F-Hmaps}
   \end{figure}



\subsection{HELICITY FLUX MAPS}
\label{Ss-Hmaps}

The helicity flux distribution estimated with \gth\ is displayed in
Figure~\ref{F-Hmaps} at several times before the flares. As
expected, the maps are strongly dominated by negative helicity
patches, the main one being located in the central intense negative
polarity N1 (see Figure~\ref{mdi}). The magnitude of helicity density
is typically of  $15 \times 10^{6} \U{Wb^2\, m^{-2}\,s^{-1}}$ at
that location. This is twice larger than the typical flux density in
the surrounding polarities (N2/P2; N3/P3) and about one order of
magnitude larger than in the southern polarities (N4/P4), where the
absolute flux densities are of the order of $2 \times 10^{6}
\U{Wb^2\, m^{-2}\,s^{-1}}$.

Even though the maps are dominated by the negative injection in N1,
one observes positive patches in the south of the AR, around the
polarities P2/N2, P3/N3 and P4/N4 (see Figure~\ref{F-Hmaps}). These
positive patches have a lower intensity but are more extended than
the one in N1 corresponding to the main helicity injection. In addition, even if
extended patches of positive helicity flux are observed, patches of negative
helicity flux are present nearby. So what is the real
helicity flux injection in this area? Are all these patches real or
spurious signals?

Methods estimating the helicity flux fail to produce fully reliable helicity flux
density maps because the helicity flux density per unit surface (as
\ga\ and \gth) is not a physically meaningful quantity. Magnetic
helicity flux  density can only be properly defined for an
elementary flux tube (\opencite{Pariat05}). The physically
meaningful helicity flux density through a surface is the helicity
flux density per unit flux, {\it i.e.} the weighted sum of helicity
injected at both footpoints of a magnetic field line. The proxy
\gth\  only represents a particular distribution of the helicity
injected in an elementary flux tube over its footpoint. To properly
determine the helicity flux injection, it is therefore necessary to
have a knowledge of the field line connectivity. Summing the
helicity injection in a connectivity domain is a way to obtain a
correct information about the helicity flux. Therefore, in order to
estimate the helicity flux in the southern polarities we need to
determine their connectivity before the flare.


\subsection{LOCALIZED HELICITY INJECTION}
\label{Ss-Hconnectivity}

TRACE images (for example see Figure~\ref{trace}, bottom panel) give
little information about the real connectivity around polarities
P2/N2, P3/N3 and P4/N4. Therefore, we use the extrapolation
(Section~\ref{S-Topology}) to estimate the connectivity of the south
polarities. We use the magnetic field model with
$\alpha=-1.2\times10^{-2}\U{Mm^{-1}}$ computed from the MDI magnetogram
at 06:23 UT, just prior to the flare. We are assuming that it
accurately represents the connectivity during the previous two days,
{\it i.e.} we suppose that extremely limited reconnection has
occurred that would have changed the connectivity. We also perform
an extrapolation with $\alpha=1.2\times10^{-2}\U{Mm^{-1}}$. We
observe that the field line connectivity is relatively insensitive
to the variation of $\alpha$. This is not surprising since field
lines anchored in the south polarities are relatively short and
the variation of $\alpha$ primarily affects field
lines with lengths of the order of $\alpha^{-1}$, as previously
discussed. Nonetheless, this connectivity conservation in spite of
the change in $\alpha$ gives us confidence on our following results.


\begin{figure} 
\centerline{\hspace*{0.02\textwidth}
               \includegraphics[width=0.80\textwidth,viewport=25 130 587 600,clip=]{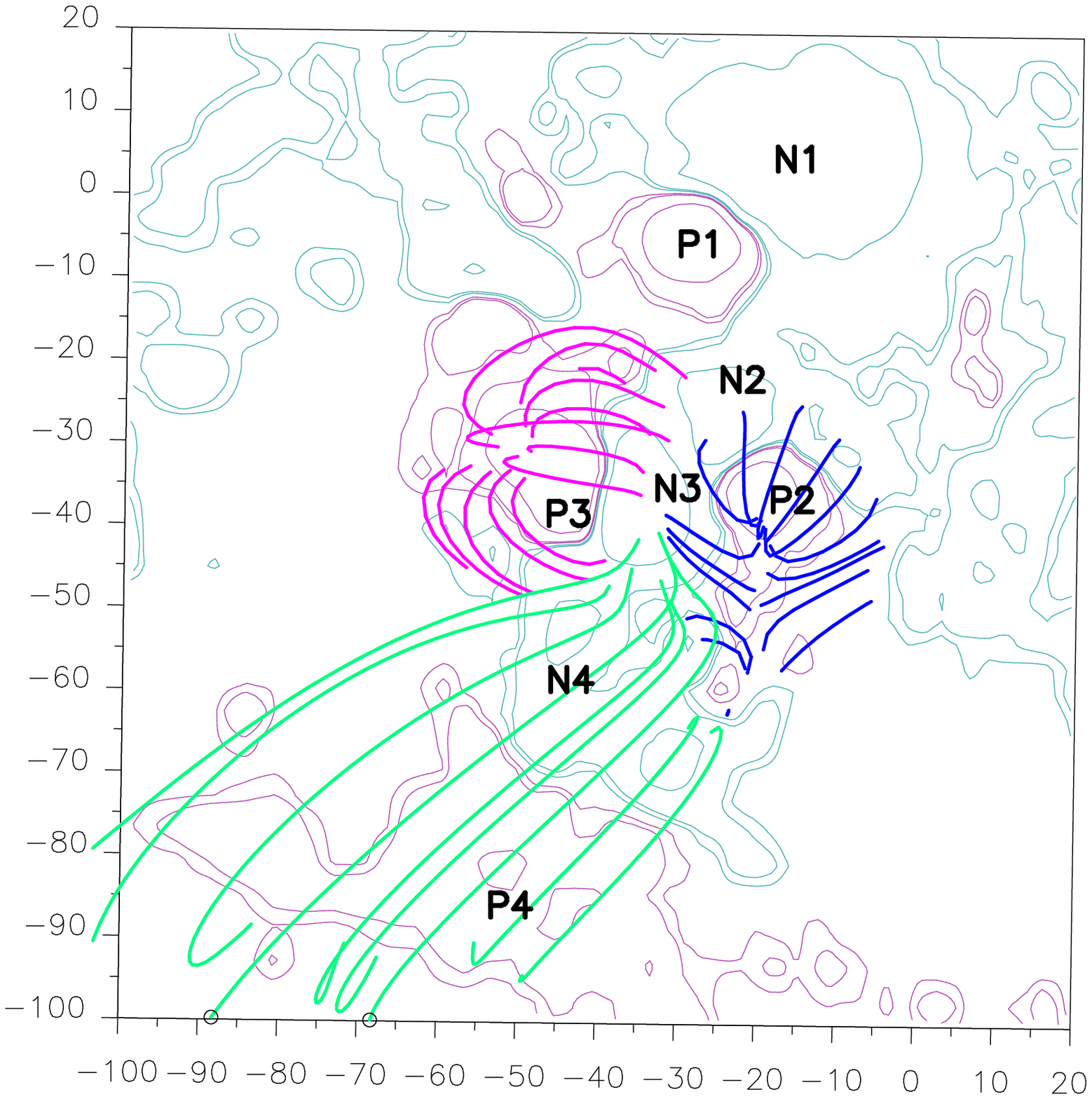}
              }     \vspace{0.05\textwidth}    
\caption{Field lines (06:23 UT) computed to show the three
connectivity domains around polarities N2/P2, N3/P3 and N4/P4. Each
set of field lines (different colors) presents a different
connectivity domain: The P2 domain is displayed with blue lines, the
P3 domain with pink lines and the P4/N4 domain with green lines.}
   \label{F-Connectivity}
   \end{figure}


\begin{figure} 

\centerline{\hspace*{0.02\textwidth}
               \includegraphics[width=0.50\textwidth,clip=]{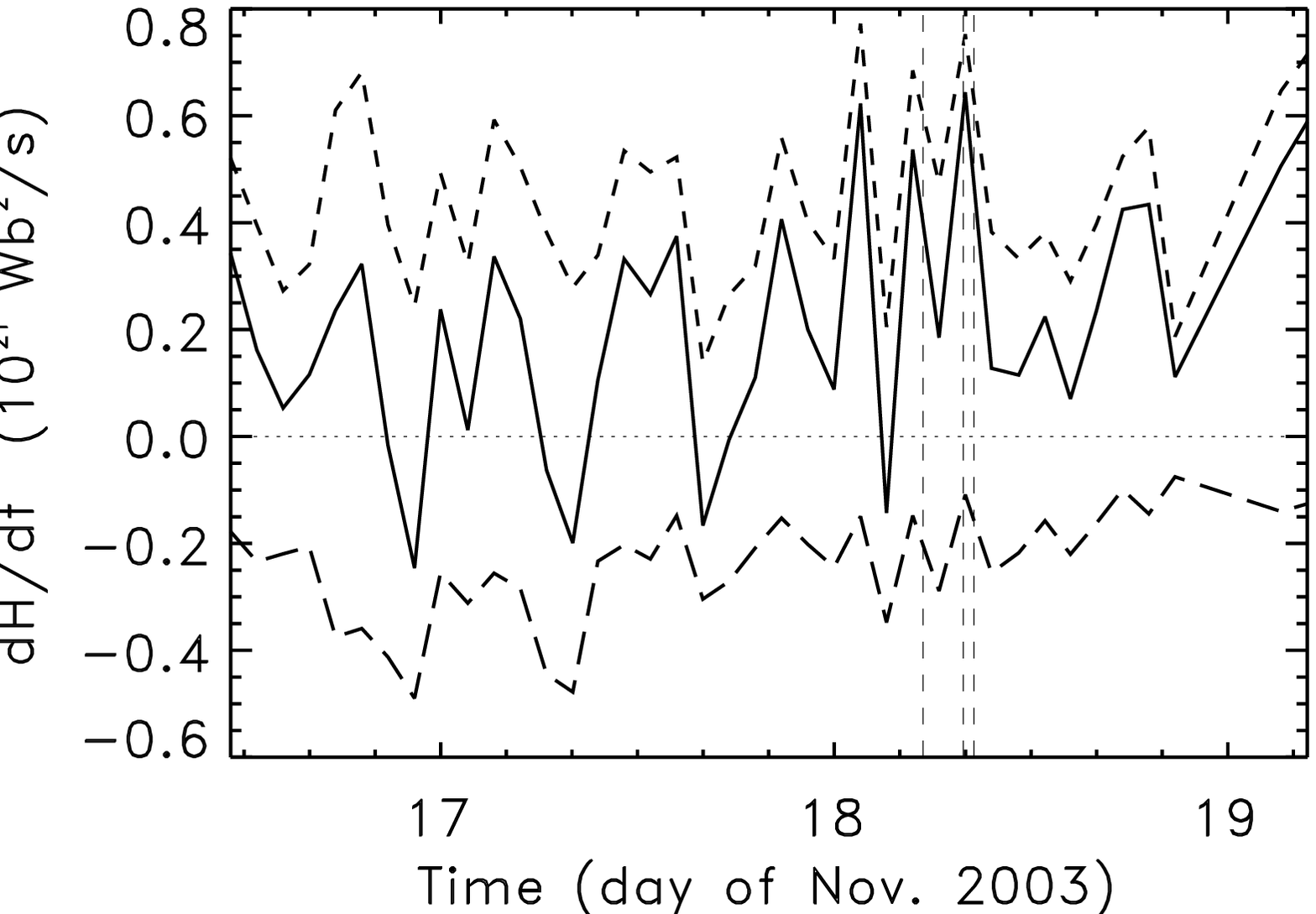}
                \includegraphics[width=0.50\textwidth,clip=]{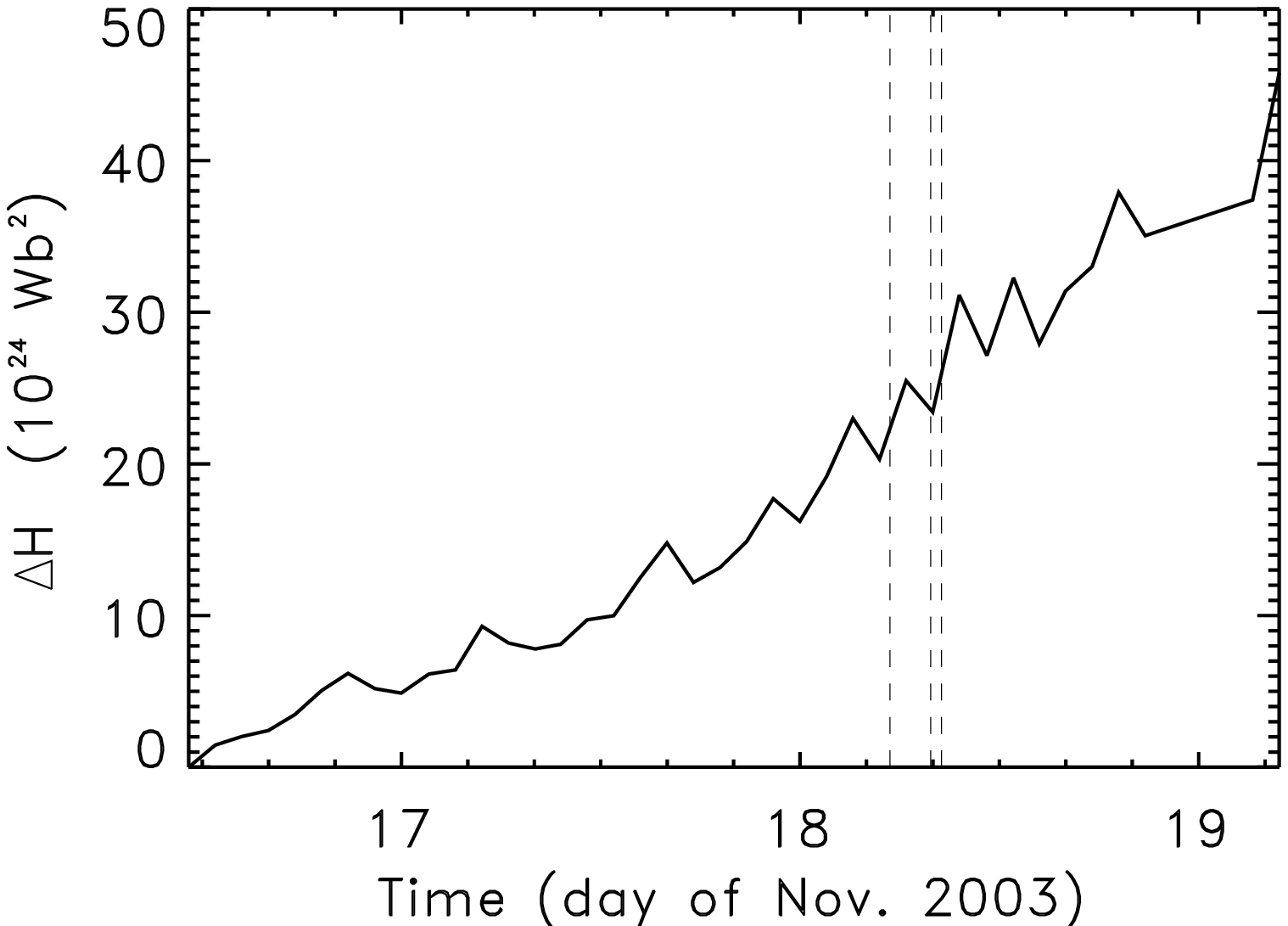}
              }
     \vspace{0.05\textwidth}    
\caption{Positive (dashed), negative (long-dashed), and total (solid) helicity fluxes (left) and accumulated
helicity (right) in N4/P4 domain of AR 10501 between 16 November
11:11 UT and 19 November 04:51 UT. The description of the figure is
the same as in Figure \ref{F-Hflux}}
   \label{F-Hflux_south}
   \end{figure}
We determine the existence of three connectivity domains around
P2/N2, P3/N3 and P4/N4 (see Figure~\ref{F-Connectivity}). Whereas P4
and N4 are simply connected, we observe that field lines originating
from N2 and N3 actually split between P2 and P3. A
(quasi-)separatrix is likely to lie in the middle of N2 and N3. The
three connectivity domains are thus the one associated to P2 (blue
lines on Figure~\ref{F-Connectivity}), the one associated with P3
(pink lines on Figure~\ref{F-Connectivity}) and the P4/N4 one (green
lines on Figure~\ref{F-Connectivity}).

Given the connectivity, we are able to estimate the helicity in each
domain. We roughly approximate these domains as rectangular sub-maps
of the MDI magnetograms, {\it i.e.} these sub-maps are delimited by eye
in order to fit the domain in the extrapolation. We do not believe
that a more sophisticated method is necessary given the uncertainty
on the extrapolation and the fact that we mostly want an estimate of
the total helicity flux in each of these domains. We nonetheless
control the amount of magnetic flux unbalance in each of these
domains. Balanced magnetic flux indeed implies that the boundaries
of our domains are well defined (in a given domain the flux should be
perfectly balanced by definition). In none of the computed domains
we find a contrast, C, larger than 0.2, with
$C=(|\Phi_+|-|\Phi_-|)/(|\Phi_+|+|\Phi_-|)$, with $|\Phi_\pm|$ the
total positive (negative) flux.

Integrating the helicity flux density in each of these connectivity
domains, we confirm the result that a positive helicity injection
occurred in the southern polarities. More precisely, we find that in
the P3 domain the helicity injection is extremely mixed. Strong
patches of both signs are present. The helicity accumulation is
first negative ($-4\times 10^{24} \U{Wb^2}$), then it becomes
positive ($9\times 10^{24} \U{Wb^2}$) between $~$ 00:00 UT and 17:35 UT
on 17 November, to return to be slightly negative after. In the P2
connectivity domain the helicity injection is more clearly positive.
However, the accumulated helicity is not larger than $5\times
10^{24} \U{Wb^2}$ during the two days preceding the flaring events.

Actually, the largest positive helicity injection is found to occur
in the third domain, P4/N4. Even though the helicity flux density is
relatively small, this is the region where a much extended patch of
positive helicity flux is observed. Even though the computed
helicity density is low, we can confirm that the helicity computations
with different values of $\omega$ and noise do not sensitively
change the results. We determine that in P4/N4 a total helicity
larger than $30 \times 10^{24} \U{Wb^2}$ was injected between 16
November 11:11 UT and 18 November 07:59 UT (see
Figure~\ref{F-Hflux_south}). Would this injection have occurred
during six days at the same rate, an accumulated helicity of the order
of $10^{26} \U{Wb^2}$ would have been injected, enough to explain
the helicity carried by the positive MC.

Studying the flux transport velocity field $\uu$, it is also
possible to understand why positive injection occurs there. Two kinds of motions in the
photospheric magnetic polarities occurred at the location of N4/P4. First,
there is a global southward displacement of P4 and N4 (see
Figure~\ref{F-Hmaps}, left panels). However, we see that this
displacement is more pronounced for P4. Second, following
isocontours of the magnetic field, we observe that N4 and P4 have a
relative converging motion. This motion cannot plausibly inject a
large amount of helicity. However, we also observe in
Figure~\ref{F-Hmaps}, that $\uu$ presents a dominant west
orientation in P4.  This reveals the fact that P4 is actually
rotating westward relatively to N4. A counterclockwise rotation of a
positive polarity around a negative polarity (injecting positive
helicity) is perfectly consistent with the helicity injection
measured in N4/P4.

Overall, the careful study of the helicity density maps allows us to
confirm that, even if AR 10501 presents a global negative helicity,
a localized positive injection occurs at the south of the main AR
spot, at the location of polarities N4/P4, precisely where an
erupting segment of the filament is rooted.

\section{Conclusions} 
  \label{S-Concl}

We have determined the global and local magnetic helicity sign
of AR 10501 on 18 November 2003, using different methods based on
the analysis of a multi-wavelength data set. We have also discussed
the association of this AR with the MC of 20 November 2003,
observed by ACE, the largest geoeffective cloud during the solar cycle 23.

AR 10501 is surrounded by a large apparent circular-shaped
single filament, which is in fact formed by several
distinct segments. The flares of  18 November 2003 were initiated by the
continuous emergence of magnetic bipoles
and by the eruption of some segments of the filament.  The
destabilization of one of the filament segments is the primary
trigger of the third flare. As discussed in the `CSHKP'
(\opencite{Carmichael64}; \opencite{Sturrock66};
\opencite{Hirayama74}, and \opencite{Kopp76}; also Forbes and Malherbe (1991))
standard solar-flare model, a filament eruption (in our
particular case, the eruption of a filament segment) due to some
magnetic instability occurs above the magnetic inversion line. As a
result, filament material moves away from the solar surface,
the pre-existing magnetic arcade stretches upward, and the condition
arises for magnetic reconnection, leading to a CME ejection and
a subsequent MC in the interplanetary medium.

Based on the multi-wavelength data set, we have found the following
results in relation to the magnetic helicity of the AR:

\smallskip

\noindent $-$ The very extended filament
within and surrounding the AR presents evidence of dextral, as well
as sinistral chirality, {\it i.e.} negative and positive magnetic
helicity.

\noindent $-$ The sunspot whorls show a left-hand twist, which
corresponds to negative magnetic helicity.

\noindent $-$ The reverse `J-shaped' flare ribbons indicate
negative helicity in the active region. However, this may be a
false indicator because the ribbons follow the magnetic inversion
line.

\noindent $-$ From the coronal magnetic field model, we have found
that the best $\alpha$ values to fit the large-scale TRACE loops are
negative. These negative $\alpha$ values mean that the magnetic
helicity is globally negative. However, at the south of the AR,
where N4/P4 lie, the value of $\alpha$ needed to locally match the
shape of loops in the `post-flare' arcade after the flare that
peaked at 08:31 UT has a tendency to be positive towards the
east and negative towards the west, in agreement with the location
of sinistral and dextral filament segments.

\noindent $-$ The computation of the magnetic helicity injection,
using G$_\theta$ maps,  indicates that the main spot in the AR is dominated
by negative helicity. However, there is a strong local positive injection
of helicity in the southern polarities (N4/P4) of the AR.

\smallskip
From the above evidences, we conclude that AR 10501 has a global
negative magnetic helicity. Despite this global negative helicity,
the helicity density maps, {\it i.e.} G$_\theta$ maps, show a strong
injection of positive magnetic helicity in the southern polarities.
Our result provides a clear example of an AR in which the
magnetic helicity sign is mixed, with simultaneous injection of both
helicity signs. Previous works (\opencite{Pevtsov97};
\opencite{Green02}) have already reported examples in which the
total helicity sign of an AR changed as it evolved because of
parasitic magnetic polarity emergence having an opposite helicity
sign to the main one. This is the first time that helicity flux
density maps bring new information about 
the local helicity injection into an active region whose sign is opposite
to the global helicity of the AR. Similar 
local injection of helicity of an opposite sign may explain the few cases of discrepancy found by
\inlinecite{Leamon04} between the helicity sign carried by MCs and
the global helicity sign of their identified solar source region.

We speculate that due to a global instability in the AR, a flux rope
with positive helicity erupted. This flux rope would contain the
filament segment with positive helicity. As discussed above, the
erupting filament has sections with opposite chiralities (sinistral
in F2a and dextral in F2b). Filaments with the same chirality can
merge, while those with different chirality cannot merge
(\opencite{Martin98}; \opencite{Rust01}; \opencite{Schmieder04};
\opencite{Aulanier06}). 
The segment F2 corresponds to two different flux tubes having the
same axial magnetic field direction in agreement with the
counterstreaming motions observed 
before the eruption. However, the two sections F2a and F2b have
opposite chiralities. Possibly, during the third flare, these
two segments interacted through magnetic reconnection and helicity
carried by both section partially canceled. The positive helicity
carried by F2a being larger, the CME and ICME induced by the
eruption transported away a net positive helicity, as measured by
ACE at 1 AU \cite{Gopalswamy05}.

Finally, our finding of a local injection of helicity with
opposite signs is potentially important for eruption models.
\inlinecite{Kusano04} developed a model based on the reconnection of
magnetic flux ropes with opposite helicity. Helicity
annihilation potentially allows the release of a larger amount of
free energy since the field can eventually relax to a state closer
to potential.

\begin{acks}
The authors thanks Dr. Pascal D\'emoulin for fruitful discussions
and suggestions. R.C. thanks the Le Centre Franco-Indien pour la
Promotion de la Recherche Avanc\'ee (CEFIPRA) for his postdoctoral
grant. This work was done in the frame of the European network
SOLAIRE. This work used the DAVE/DAVE4VM codes written and developed
by the Naval Research Laboratory. E.P. wishes to thank Peter Schuck
for providing the DAVE/DAVE4VM code and for useful discussions. The
work of E.P. was supported, in part, by the NASA HTP and SR$\&$T
programs. C.H.M. thanks the Argentinean grants: UBACyT X127 and PICT
03-33370 (ANPCyT). C.H.M. is a member of the Carrera del
Investigador Cient\'{i}fico, CONICET. We acknowledge the use of
TRACE data. MDI data are a courtesy of SOHO/MDI
consortium. SOHO is a project of international cooperation between
ESA and NASA. B.S and C.H.M have started this work in the frame
of the ISSI workshop chaired by Dr. Consuelo Cid (2008-2010). We also thank the
anonymous referee for helpful and constructive comments.
\end{acks}


\mbox{}~\\
\bibliographystyle{spr-mp-sola}
\IfFileExists{\jobname.bbl}{} {\typeout{}
\typeout{***************************************************************}
\typeout{***************************************************************}
\typeout{** Please run "bibtex \jobname" to obtain the bibliography}
\typeout{** and re-run "latex \jobname" twice to fix references}
\typeout{***************************************************************}
\typeout{***************************************************************}
\typeout{}}

\end{article}
\end{document}